\RequirePackage{fix-cm}
\RequirePackage[figuresright]{rotating}
\documentclass[twocolumn]{svjour3}
\smartqed
\usepackage{cite}
\usepackage{graphicx}
\usepackage{textcomp}
\usepackage{mathtools}
\usepackage{afterpage}
\usepackage{tabularx}
\usepackage{hyperref}
\usepackage{xcolor}

\begin{document}

\title{Fabrication of Optical Nanofibre-Based Cavities using Focussed Ion-Beam Milling - A Review}

\author{Priscila Romagnoli \and
        Maki Maeda \and 
        Jonathan M. Ward \and
        Viet Giang Truong \and
        S\'ile {Nic Chormaic}}

\institute{Light-Matter Interactions for Quantum Technologies Unit,
            Okinawa Institute of Science and Technology Graduate University,
            \emph{Onna, Okinawa 904-0495, Japan}
            \\\ 
            \\\ \email{priscila.romagnoli@oist.jp, sile.nicchormaic@oist.jp}}

\date{Received: date / Accepted: date}
\maketitle
\begin{abstract} \label{abs}


Nanofibre-based optical cavities are particularly useful for quantum optics application,  such as the development of integrated single-photon sources, and for studying fundamental light-matter interactions in cavity quantum electrodynamics (cQED).
Although several techniques have been used to produce nanofibre-based optical cavities, focussed ion beam (FIB) milling is becoming popular; it can be used for the fabrication of complex structures directly in the nanofibre.
This technique uses a highly accelerated ion beam to remove atoms from the target material with high resolution.
However, it is challenging to mill insulating materials with highly-curved structures and large aspect ratios, such as silica nanofibres, due to charge accumulation in the material that leads to mechanical vibrations and misalignment issues.
In this article, we  highlight the main features of nanofibres and briefly review cQED with nanofibre-based optical cavities. An overview of the milling process is given with a summary of different FIB milled devices and their applications.
Finally, we present our technique to produce nanofibre cavities by FIB milling.
To overcome the aforementioned challenges, we present a specially designed base plate with an indium tin oxide (ITO)-coated Si substrate and outline our procedure, which improves stability during milling and increases repeatability.


\end{abstract}
\section{Introduction} \label{intro}

Reducing the diameter of optical fibres to the same order of magnitude as the wavelength of the guided light can lead to some unique properties.
Not only does transverse confinement of the guided modes increase, but an intense evanescent field also extends from the boundary of the fibre
\cite{Birks_JLT_1992, Tong_N_2003, Tong_OE_2004, Brambilla_AOP_2009, Tong_Sumetsky_2010} into its surroundings. In fact, in some cases, the amount of energy in the evanescent field exceeds what remains inside the fibre, so that the light is essentially guided outside the fibre. The ease with which these optical microfibres (OMF) or optical nanofibres (ONF) can be integrated into experimental setups means that they offer many advantages for applications in areas as diverse as optical sensing \cite{Villatoro_OE_2005, Lou_S_2014, Beugnot_NC_2014, Brambilla_OFT_2010, Tong_S_2018}, 
atomic physics \cite{Sague_PRL_2007, Vetsch_PRL_2010, Hendrickson_PRL_2010, Jones_JOSAB_2014, Kien_PRA_2017, Kien_PRA_2018}, photonics \cite{Aoki_N_2006, Shoppova_APL_2007, Ward_LPR_2011, Lei_OE_2017} and quantum optics \cite{Nayak_NJP_2008, Ruddell_O_2017, Solano_Adv_2017, Kato_NC_2019, Araneda_NP_2019}.

Extensive studies have shown that confining photons in optical cavities can increase the strength of interactions between the photon emitter and the guided mode of the cavity \cite{Haroche_PT_1989, Kimble_IOP_1998, Haroche_CQE_1995, Mabuchi_S_2002, Vahala_N_2003, Kimble_N_2008, Reiserer_RMP_2015}.
Therefore, the combination of optical nanofibres with optical cavities is an ideal platform for studying cQED because optical cavities permit long photon storage times and small mode volumes in a device that is directly integrated into an optical fibre \cite{Kien_PRA_2009, Nayak_OE_2011, Wuttke_OL_2012, Nayak_OE_2013, Yalla_PRL_2014, Daly_SPIE_2015, Kato_PRL_2015, Nayak_JO_2018}.

In the literature, there are several different techniques describing how to produce cavities directly in optical nanofibres \cite{Nayak_OE_2013, Nayak_OL_2014, Yalla_PRL_2014, Daly_SPIE_2015, Daly_OE_2016, Daly_PHD_2017, Keloth_OL_2017, Li_APL_2017}.
One method that is proving particularly successful is the use of FIB milling to fabricate Bragg mirrors for fibre-based optical cavities \cite{Nayak_OE_2011, Schell_SR_2015, Li_APL_2017, Takashima_OE_2019}. 
This technique relies on a highly focussed ion beam to cut patterns into the fibre with high resolution.
However, FIB milling on highly curved structures with large aspect ratios, such as silica nanofibres, is difficult in practice due to mechanical vibrations and the resulting misalignment due to charge accumulation on the device.
To improve repeatability and quality of the fabrication process, methods must be developed to mitigate these challenges.

In this article, we present a brief review of techniques used for FIB milling of nanofibre-based cavities and the progress that has been achieved to date.
In Section \ref{nanofibre_cavities} we review important properties of optical nanofibres and introduce cQED for fibre-based platforms.
Section \ref{fib} discusses the FIB milling technique and reviews previous work on milling optical fibres, including micro- and nanofibres, with emphasis on fibre-based cavities.
In Section \ref{experiment} we present our technique to fabricate nanofibre-based optical cavities using FIB milling, without the need to coat the fibre with a conductive layer.
Finally, in Section \ref{conclusions}, conclusions and outlook for this technique are presented.
\section{Optical Nanofibres and Cavity Quantum Electrodynamics} \label{nanofibre_cavities}

Unique properties of optical nanofibres combined with the photon confinement provided by optical cavities have led to a number of interesting studies in quantum optics and demonstrations of efficient, integrated single-photon sources \cite{Yalla_PRL_2012, Nayak_OE_2007,  Kien_PRA_2009, Kato_PRL_2015}.
In this section we briefly discuss the main optical properties of nanofibres and how they are fabricated to maintain strict parameters needed for low-loss propagation, thereby meeting the requirements for cQED with nanofibre-based cavities.

\subsection{Optical Nanofibres - Fabrication and Mode Propagation} \label{nanofibre}

Optical fibres with diameters less than the wavelength of guided light ($d\leq\lambda$) \cite{Birks_JLT_1992, Tong_N_2003, Tong_OE_2004, Ward_RSI_2006, Brambilla_AOP_2009, Tong_Sumetsky_2010, Ward_RSI_2014} can result in an evanescent light field that  extends from the boundary of the fibre and interacts with the surrounding medium \cite{Tong_OE_2004, Kien_OC_2004}.
As a result, the guided mode is altered by the properties of the medium \cite{Patnaik_PRA_2002}.
In perhaps the most extreme example, ultrathin silica fibres with diameters as small as 50~nm and with lengths of tens of millimetres, were reported by Tong et al. \cite{Tong_N_2003}.
These nanofibres or nanowires were fabricated using a unique flame-heated fibre drawing method, in which a conventional fibre was tapered to micrometre size, with a sapphire tip placed at the end of the taper.
The sapphire tip was heated by the flame and used to draw out the nanowire.
The authors coupled light into the nanowire via the evanescent field from a tapered optical fibre attached to the nanowire by van der Waals forces.
Single-mode light guiding with low loss (below 0.1~dB~mm$^{-1}$) at a wavelength of 633~nm was achieved for a 450~nm diameter wire and also at a wavelength of 1550~nm for a 1100~nm diameter wire.
This example demonstrates the capabilities of light guidance in optical fibres that are pushed to their physical limits. These unique properties have been exploited for a large number of applications, such as nonlinear optics \cite{LeonSaval_OE_2004, Kumar_NJP_2015}, cold atoms \cite{Sague_PRL_2007, Nayak_NJP_2008, Morrissey_RSI_2009}, particle manipulation, \cite{Brambilla_OL_2007, Skelton_JQSRT_2012, Daly_NJP_2014, Daly_SPIE_2015, Daly_OE_2016} and sensors \cite{Villatoro_OE_2005, Lou_S_2014, Beugnot_NC_2014, Brambilla_OFT_2010, Tong_S_2018} and several reviews related to their fabrication and uses are available \cite{Brambilla_AOP_2009, Brambilla_OFT_2010, Brambilla_JO_2010, Tong_OC_2012, Morrissey_S_2013, Lou_S_2014, Ward_RSI_2014, Nieddu_JO_2016, Solano_Adv_2017, Tong_S_2018, Nayak_JO_2018}.

\begin{figure}
    \centering
    \includegraphics[scale=0.4]{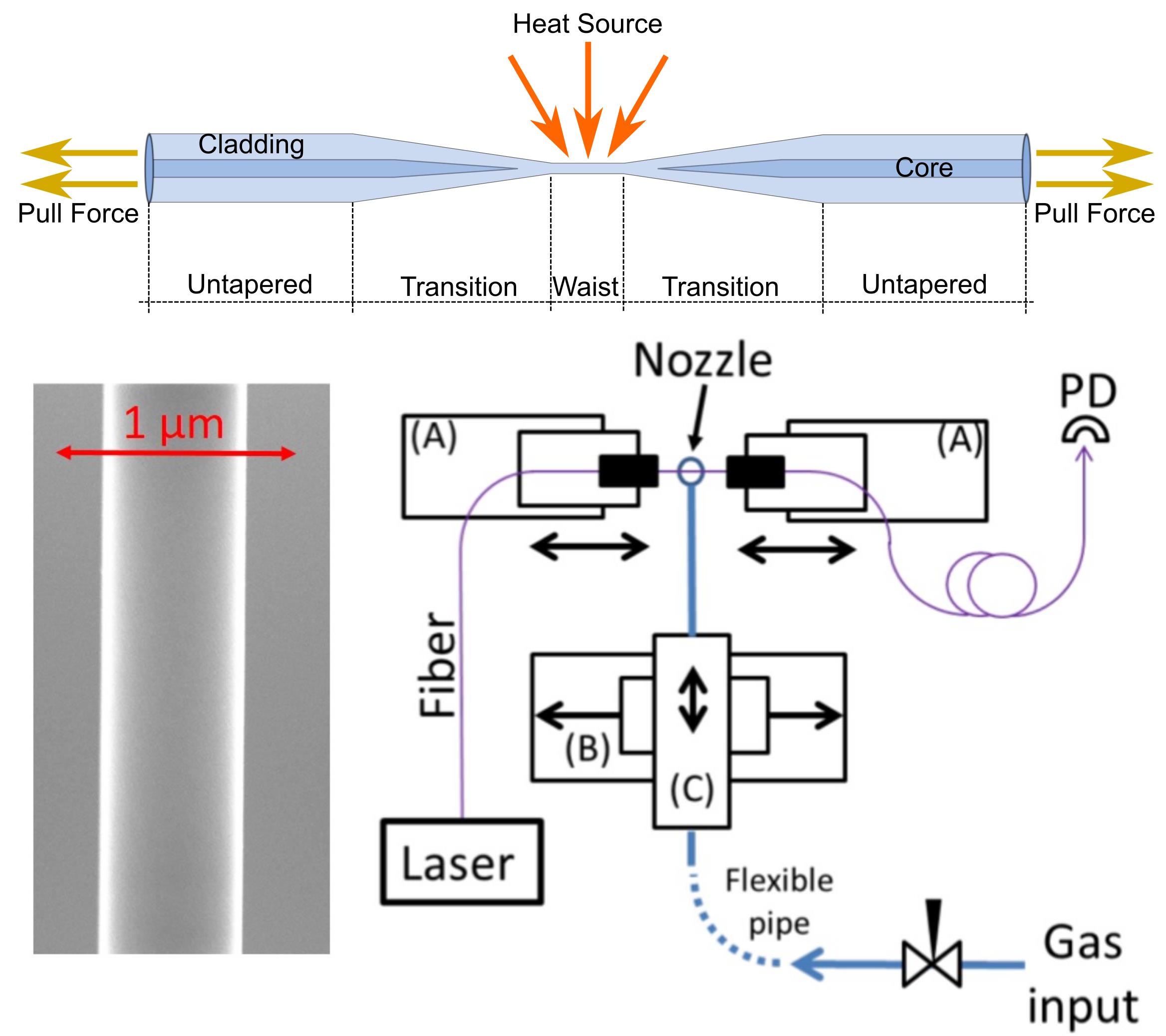}
    \caption{Top: Schematic of a tapered optical fibre. 
    The optical nanofibre is the ultrathin waist region.
    Bottom left: SEM image of an optical nanofibre.
    Bottom right: Schematic of a typical fibre pulling rig reprinted from \cite{Ward_RSI_2014}, with the permission of AIP Publishing.} \label{fig_s2_fnanofibre}
\end{figure}

Optical nanofibres are typically made by tapering a section of commercial optical fibre via heating and stretching \cite{Brambilla_OE_2004, Clohessy_EL_2005, Ward_RSI_2006, Ward_RSI_2014, Lee_CAP_2019} in a fibre pulling rig, with either a stationary or moving (brushing) heat source.
Different heat sources may be used, depending on the compound and melting point of the glass \cite{Brambilla_OFT_2010}, such as a CO$_{2}$ laser \cite{Sumetsky_OE_2004}, resistance heating \cite{Magi_OE_2007} or a graphite microheater \cite{Brambilla_EL_2005}.
After fabrication, a biconical, tapered fibre is formed, i.e., it has a down taper, transitioning from the initial diameter to a thin waist.
The waist is followed by an up taper returning to its initial diameter.
It is the thin waist region that is referred to as an \textit{optical micro- or nanofibre}.
A schematic of a tapered optical fibre, a scanning electron microscope (SEM) image of a nanofibre with a diameter of 690.7~nm, and a schematic of a typical fibre pulling rig are shown in Fig.~\ref{fig_s2_fnanofibre}.  

In this setup, the fibre is fixed on stages (referred to as A) that pull, while heating is done by an oscillating hydrogen/oxygen flame coming from a nozzle fixed on B and C.
A laser is launched into the fibre and transmission during  tapering is monitored using a photodiode.
To obtain high transmission, the adiabatic criterion has to be satisfied, i.e., the taper angle in the transition region must be small enough to avoid coupling of the guided mode to undesired higher order modes or cladding modes \cite{Petcu-Colan_JNOPM_2011}.

The shape of the taper (indicated as the transition region in Fig.~\ref{fig_s2_fnanofibre}) is determined by the size of the heated part of the fibre (the hotzone). For a fixed size of hotzone, the taper profile follows an exponential function and is suitably adiabatic for the fundamental mode.
However, linear and more complex taper shapes are also possible by varying the size of the hotzone during the pull.
Linear taper transitions are especially useful for achieving low-loss, high-order mode propagation \cite{Petcu-Colan_JNOPM_2011, Ravets_OE_2013, Ward_RSI_2014}.
Ward et al. \cite{Ward_RSI_2014} demonstrated that the flame-brushing technique can produce micro- and nanofibres with high transmission ($\sim99\%$) for the fundamental mode in single-mode fibre tapered to a diameter of  $\sim$~1.1~$\mu$m. For the first few higher order modes, that is the linearly polarised $LP_{11}$ group, transmission was $\sim95\%$ in a few-mode fibre with a diameter of 700~nm at a wavelength of 780~nm.


When making optical nanofibres, a crucial consideration for efficient low-loss light guiding is the refractive index difference between the fibre material and the surrounding medium. 
A conventional (step-index) optical fibre has a small difference between the refractive index of its cladding and core, usually on the order of $n_{cladding}-n_{core}\sim0.01$.
As a result, light guidance is considered to be in the weakly guiding regime.
When a conventional optical fibre is sufficiently tapered, the core, at the thinnest part, effectively vanishes and the cladding acts as a core, while the surrounding medium (which can be vacuum, air, or liquid) acts as a cladding \cite{Kien_OC_2004,Tong_OC_2012}.
This leads to a large difference between the refractive indices of the cladding and core.
For example, for a silica nanofibre surrounded by air, $n_{cladding}-n_{core}\sim0.45$ and light is guided in the strongly guiding regime.
The description of guided modes in the weakly guiding regime is simplified by approximating them as $LP_{lm}$ modes with the same propagation constant; in other words, the modes are degenerate for a given $l$.
The parameters $l$ and $m$ are related, respectively, to the electric field intensity distribution in the azimuthal and radial directions. 
However, in the strongly guiding regime, this approximation is no longer valid and the description of the guided modes separates them into transverse electric ($TE$), transverse magnetic ($TM$), and hybrid modes ($HE$) \cite{Snyder_Love_1983, Tong_OE_2004}.

Le Kien et al. \cite{Kien_OC_2004} calculated the field intensity distribution for the fundamental, $HE_{11}$, mode of an ultrathin fibre and showed a large discontinuity of the field distribution at the cladding/core (vacuum/silica) boundary due to high contrast between the refractive indices ($n_{cladding}-n_{core}$).
This means that, at the interface, a significant part of the guided light is in the form of an evanescent field.
This  evanescent field is one of the main advantages of using optical nanofibres as tools in quantum optics and other areas of research.
The medium surrounding the nanofibre can interact with the guided light via the evanescent component.
With this in mind, the spontaneous emission of a caesium atom near a nanofibre was studied by Le Kien et al. \cite{Kien_PRA_2005}.
The authors showed that confinement of the nanofibre-guided modes can change the decay rate of the caesium atom.
Coupling between the emission from an atom on the surface of a nanofibre, with a diameter of ~200 nm, and the guided mode was studied and a coupling efficiency of 28$\%$ was theoretically predicted.

Later experiments confirmed these theoretical predictions \cite{Nayak_OE_2007, Nayak_NJP_2008}.
For example, Nayak et al. \cite{Nayak_OE_2007} manipulated and probed the fluorescence of caesium atoms using nanofibres.
The authors employed a magneto-optical trap (MOT) to confine and cool  caesium atoms around a nanofibre with a diameter of 400~nm.
With an effective number of atoms of ~5, they showed an average coupling efficiency of spontaneous emission into the guided mode of $\sim$~6\%.
The following year, a similar setup was used by Nayak et al. \cite{Nayak_NJP_2008} to detect a single caesium atom and coupling of the atom's spontaneous emission into the modes of the nanofibre.

Aside from neutral atoms, similar experiments can be performed using a variety of single-photon emitters such as quantum dots \cite{Srinivasan_APL_2007, Fujiwara_NL_2011, Yalla_PRL_2012}, diamond nanocrystals with nitrogen vacancies \cite{Schroder_OE_2012, Liebermeister_APL_2014}, and hexagonal boron nitride \cite{Schell_ACSP_2017}. Channelling efficiency of fluorescence from quantum dots into nanofibre-guided modes was demonstrated by Yalla et al. \cite{Yalla_PRL_2012}.
The number of photons in the guided and radiation modes was measured and a channelling efficiency up to 22.0~$\pm$~4.8$\%$ was obtained for a 350~nm nanofibre and an emission wavelength of 780~nm, thus demonstrating a fibre-coupled single-photon source with a bright output in the range of 10-100 of kilocounts/second.

\subsection{Nanofibre-Based Optical Cavities for Cavity Quantum Electrodynamics} \label{cavities}

By placing a quantum emitter in an optical cavity, confinement of photons by the cavity is used in essentially two ways.
First, photons can interact with the quantum emitter multiple times, and second, light resonant in the cavity adds constructively to create high-intensity intracavity fields.
By combining an optical cavity with the small mode volumes that nanofibres can offer, it is possible to further concentrate the mode field and improve light-matter interaction rates \cite{Vahala_N_2003}.
In this subsection, the concept of cQED is introduced, with a focus on nanofibre-based cavities.

\subsubsection{Cavity Quantum Electrodynamics} \label{cavities_ov}

The spontaneous emission of an atom is not a fixed property, but rather is the result of coupling between the atom and the vacuum field.
An optical cavity can modify vacuum field fluctuations, thereby altering the properties of spontaneous emission from the atom, such as the decay rate and transition energy \cite{Haroche_PT_1989, Hood_PRL_1998, Yamamoto_SLI_1999, Gerry_Knight_QO_2005}.
The study of this phenomenon is known as cQED.

Let us consider a single-mode optical cavity coupled with an atom (assumed to be initially in the excited state of a dipole transition) and that the cavity's mode is resonant with the dipole transition.
The atom's spontaneous emission and the mode field of the cavity will couple and the energy in the system will oscillate at the vacuum Rabi frequency \cite{Vahala_N_2003}.
However, the Rabi oscillations are limited by the cavity's finite photon lifetime, i.e., energy loss from the system.
A cavity that is typically used is the Fabry-P\'erot (FP) type, where confinement of the photon is provided by two mirrors \cite{Vahala_N_2003}.
In order to quantify the efficiency of light-matter interactions, it is useful to define the cooperativity parameter, $C$, in terms of the atom-photon coupling rate, $g$, where $2g$ is the vacuum Rabi frequency, the cavity decay rate, $\kappa$, and the atomic spontaneous emission rate, $\gamma$, \cite{Kimble_IOP_1998, Nayak_JO_2018}:

\begin{equation} \label{eq_c}
C = \frac{(2g)^{2}}{\kappa \gamma} = \frac{3 Q \lambda^{3}}{4 \pi^{2} V},
\end{equation}
where
\begin{equation} \label{eq_g0}
g = \sqrt{\frac{\mu^{2} \omega_{c}}{2 \hbar \epsilon_{0} V}}.
\end{equation}
$Q$ is the quality factor of the cavity mode, $\lambda$ is the resonance wavelength, $V$ is the cavity mode volume, $\mu$ is the transition dipole moment, $\omega_{c}$ is the atomic transition frequency at $\lambda$, $\hbar$ is the reduced Planck constant and $\epsilon_{0}$ is the permittivity of free space.
An optical cavity can be characterised by the quality factor, $Q$, and the finesse, $F$.
$Q$ quantifies cavity losses and the temporal confinement of photons in the cavity, where $Q={\lambda}/{\Delta \lambda}$ and $\Delta\lambda$ is the full-width-half-maximum (FWHM) at $\lambda$.
$F$ defines the resolution capability of the optical cavity, where $F={FSR}/{\Delta\lambda}$ and $FSR$ is the free-spectral-range.

For strong interactions between an atom and a photon, it is required to have $C \gg 1$ \cite{Keloth_OL_2017}.
However, even for $C \gg 1$ there are two coupling regimes that can be used to classify cQED.
The first regime occurs when $\kappa > 2g, \gamma$ and is termed weak coupling \cite{Fox_QO_2006, Keloth_OL_2017} or the Purcell regime, where the atom-cavity interaction is slower than the dissipation \cite{Yamamoto_SLI_1999}.
However, there is still an enhancement of spontaneous emission of the atom compared to free space and it can be defined by the Purcell factor, $P$.
For the weak coupling regime, the Purcell factor can be defined as $P \approx C$ \cite{Purcell_PR_1946, Yalla_PRL_2014}.

The second regime is when $2g > \gamma,\kappa$ and is termed strong coupling \cite{Fox_QO_2006, Keloth_OL_2017}.
This occurs when a cavity can support, even if only briefly, Rabi oscillations in spite of the energy leaking out \cite{Vahala_N_2003}.
In other words, the atom-cavity interaction is faster than the dissipation and the emitted photon remains in the cavity long enough to have a high probability of being reabsorbed by the atom \cite{Haroche_PT_1989, Yamamoto_SLI_1999}.

From equations \ref{eq_c} and \ref{eq_g0} it is clear that a small cavity volume, combined with a high quality factor, can increase cooperativity.
If we consider a cavity formed by a nanofibre between two mirrors  \cite{Kato_PRL_2015, Li_APL_2017}, the separation between the mirrors and the diameter of the nanofibre can be engineered so as to yield small mode volumes with relatively large $Q$ factors.
Additionally, in 2009, Le Kien et al. \cite{Kien_PRA_2009} showed that for nanofibre-based cavities, $C$ is independent of the cavity length, and even for moderate-finesse cavities, high cooperativity can be achieved.
The particulars of nanofibre cavities make them promising tools for optimising light-matter interactions, as briefly discussed in the following section.

\subsubsection{Nanofibre-Based Optical Cavities} \label{nanocavities}

\begin{figure}
    \centering
    \includegraphics[scale=0.35]{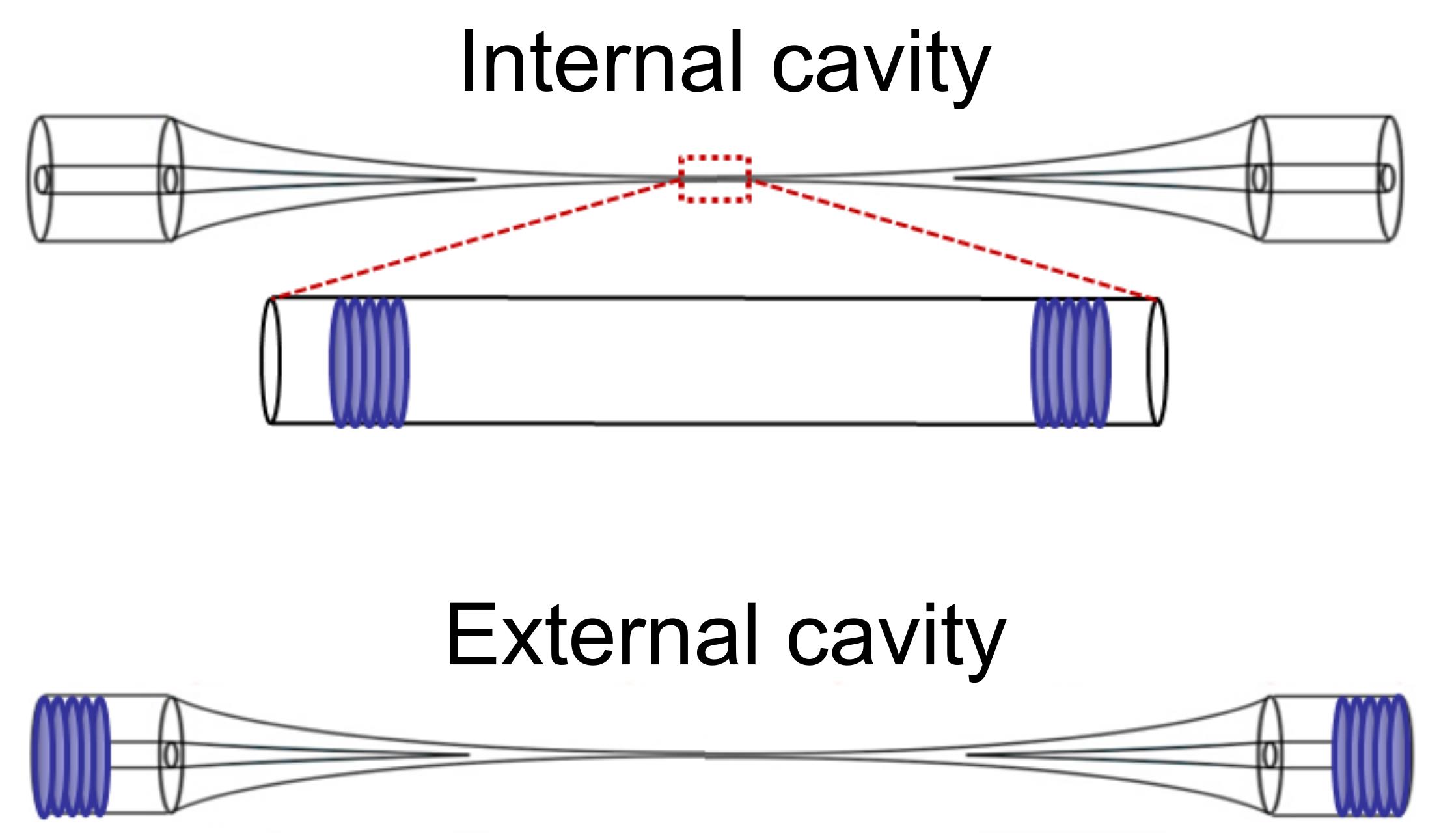}
    \caption{Top: internal cavity scheme. The mirrors are fabricated directly on the nanofibre region.
    Bottom: external cavity scheme. The mirrors are fabricated outside the tapered region.} \label{fig_s2_nanocavity}
\end{figure}

Nanofibre-based optical cavities can be viewed as inline optical cavities, since the structure is built into a single fibre rather than creating a cavity from the reflective ends of two separate fibres.
In-line Fabry-P\'erot type cavities can be produced by two different approaches:
(i) the mirrors are fabricated directly on the nanofibre region and are designated hereafter as internal cavities (see Fig.~\ref{fig_s2_nanocavity} top) or
(ii) the mirrors are fabricated outside the nanofibre region, but still on or within the fibre. We refer to these as external cavities (see Fig.~\ref{fig_s2_nanocavity} botom) \cite{Nayak_JO_2018}.
In both approaches, the most common method to produce the cavity mirrors is to fabricate fibre Bragg gratings (FBGs) by generating strong, permanent modulation of the refractive index in the fibre.

Le Kien et al. \cite{Kien_PRA_2009} developed a theoretical formalism to understand dynamics of the interaction between a nanofibre-based cavity created by FBGs and the spontaneous emission of a single, nearby atom.
They showed that the atom-photon coupling rate for a nanofibre-based cavity, $g_{nfc}$, becomes \cite{Kien_PRA_2009, Nayak_JO_2018}:

\begin{equation} \label{eq_gf}
g_{nfc} = \sqrt{\frac{c \gamma \eta}{L},}
\end{equation}
where $c$ is the speed of light in vacuum, $\eta$ is the channelling efficiency near a nanofibre without a cavity, and $L$ is the cavity length.
In  equation (\ref{eq_gf}), $\gamma$ refers to the rate near a nanofibre without a cavity and contains the contribution from the guided ($\gamma_{g}$) and radiation ($\gamma_{r}$) parts, with $\gamma = \gamma_{g} + \gamma_{r}$, and $\eta = \gamma_{g} / \gamma$.
Therefore, the cooperativity parameter for nanofibre cavities ($C_{nfc}$) can be rewritten as \cite{Nayak_JO_2018}:

\begin{equation} \label{eq_cf}
C_{nfc} = \frac{4}{\pi} \eta P F.
\end{equation}

Notice that $C_{nfc}$ does not depend on cavity length.
Ref. \cite{Kien_PRA_2009} concluded that, even with a moderate finesse cavity, coupling between the atom and nanofibre guided mode could be enhanced.
For the weak coupling regime, they found that with a 200~nm diameter fibre and a cavity finesse of about 30, up to 94$\%$ of the spontaneous emission from the atom could be coupled into the nanofibre cavity.
Additionally, it was shown that strong coupling could be achieved for a moderate finesse of 30 and a reasonably long cavity of 10~cm to 1~m.

Experimentally, an external FBG nanofibre cavity was demonstrated by Wuttke et al. \cite{Wuttke_OL_2012} in 2012.
The cavity was in the strong coupling regime with a finesse of 86 and was resonant with the D$_{2}$  line of atomic caesium.
Similarly, in 2015, Kato et al. \cite{Kato_PRL_2015} observed strong coupling between the same type of cavity (an external FBG nanofibre cavity) and a single trapped caesium atom.
The coupling regime was tuned by temperature, since one of the FBGs was designed to have the edge of its reflection band at the D$_{2}$ caesium emission wavelength.
The nanofibre had a 400~nm diameter and the cavity length and finesse were, respectively, 33~cm and 40.

In 2013, Nayak et al. \cite{Nayak_OE_2013} experimentally demonstrated a fabrication technique to create  \textit{internal} FBG nanofibre cavities.
The authors could produce thousands of periodic nanocraters on the nanofibre using a femtosecond (fs) laser ablation technique. For nanofibres with diameters ranging from 450-650~nm, crater diameters down to 95~nm were formed.
This structure can act as a one-dimensional photonic crystal (PhC) and provides strong confinement of the electric field in both the transverse and longitudinal directions.

Thereafter, the fs laser ablation technique continued to produce high-quality cavities \cite{Nayak_OL_2014, Keloth_OL_2017}.
Keloth et al. \cite{Keloth_OL_2017} reported on a 1.2~cm cavity produced with this technique that supported both the weak and strong coupling regimes.
The nanofibre was $\sim$1.7~cm long and had a diameter of $\sim$500~nm.
The authors observed four modes in the cavity and showed that the modes with finesse between 200-400 were suitable to achieve strong coupling, while other modes were only suitable for the weak coupling regime.

A different approach to produce nanofibre cavities in the weak coupling regime was presented by Yalla et al. \cite{Yalla_PRL_2014}. 
The authors created a composite photonic crystal cavity, formed by placing a grating with a defect in contact with the nanofibre.
The grating was fabricated by electron beam lithography on a silica substrate.
The structure was polarisation-dependent and the cavity quality factors measured for the $x$ and $y$ polarisations were 1410 and 2590, respectively.
They also observed enhancement 
of the spontaneous emission rate from colloidal quantum dots into the nanofibre-guided modes.

In the following section, we discuss important aspects of the FIB milling technique \cite{Nayak_OE_2013, Nayak_OL_2014, Yalla_PRL_2014, Keloth_OL_2017, Li_APL_2017} and highlight some FIB milled fibre-based devices and their applications.  Finally a review of fibre-based optical cavities made by FIB is presented.  
\section{Focussed Ion Beam Milling} \label{fib}

The focussed ion beam (FIB) technique  relies on different interactions between energetic ions and atoms from the surface of a target material. It is a versatile tool widely used for imaging, milling, and deposition and has found uses in the semiconductor industry, material science, and even biology \cite{Reyntjens_JMM_2001, Tseng_JMM_2004, Tseng_S_2005, Keskinbora_FIB_2019}. Before we proceed, let us briefly outline the operation and main components of the machine itself.

\subsection{Overview of the FIB Technique} \label{fib_overview}

Ions are heavier than electrons; therefore, at a given acceleration, they can have a higher energy when they strike a target material \cite{Tseng_JMM_2004}.
Various interactions can occur when an energetic ion collides with a material, such as sputtering, deposition, redeposition, backscattering, implantation, swelling, and nuclear reaction \cite{Tseng_JMM_2004}.
The dominant interactions for FIB milling are sputtering and redeposition \cite{Tseng_S_2005}.
In the FIB technique, an ion beam is accelerated and collides with a target material.
For milling, an energetic ion collision can transfer enough energy to overcome the binding energy of chemically bound atoms on the material surface \cite{Martelli_OE_2007}. 
The result is a sputtering interaction, in which atoms are ejected from the material \cite{Sigmund_PR_1969}.  
However, a redeposition interaction also can take place during the milling process.
This occurs because the sputtered atoms (in a gas phase) are not in thermodynamic equilibrium and can condense back into a solid if they collide with any nearby surface \cite{Tseng_S_2005}.
This means that the sputtered material can redeposit in the milled area; therefore, it is necessary to control sputtering and redeposition in order to achieve precise milling \cite{Tseng_JMM_2004}.

\begin{figure}
    \centering
    \includegraphics[scale=0.7]{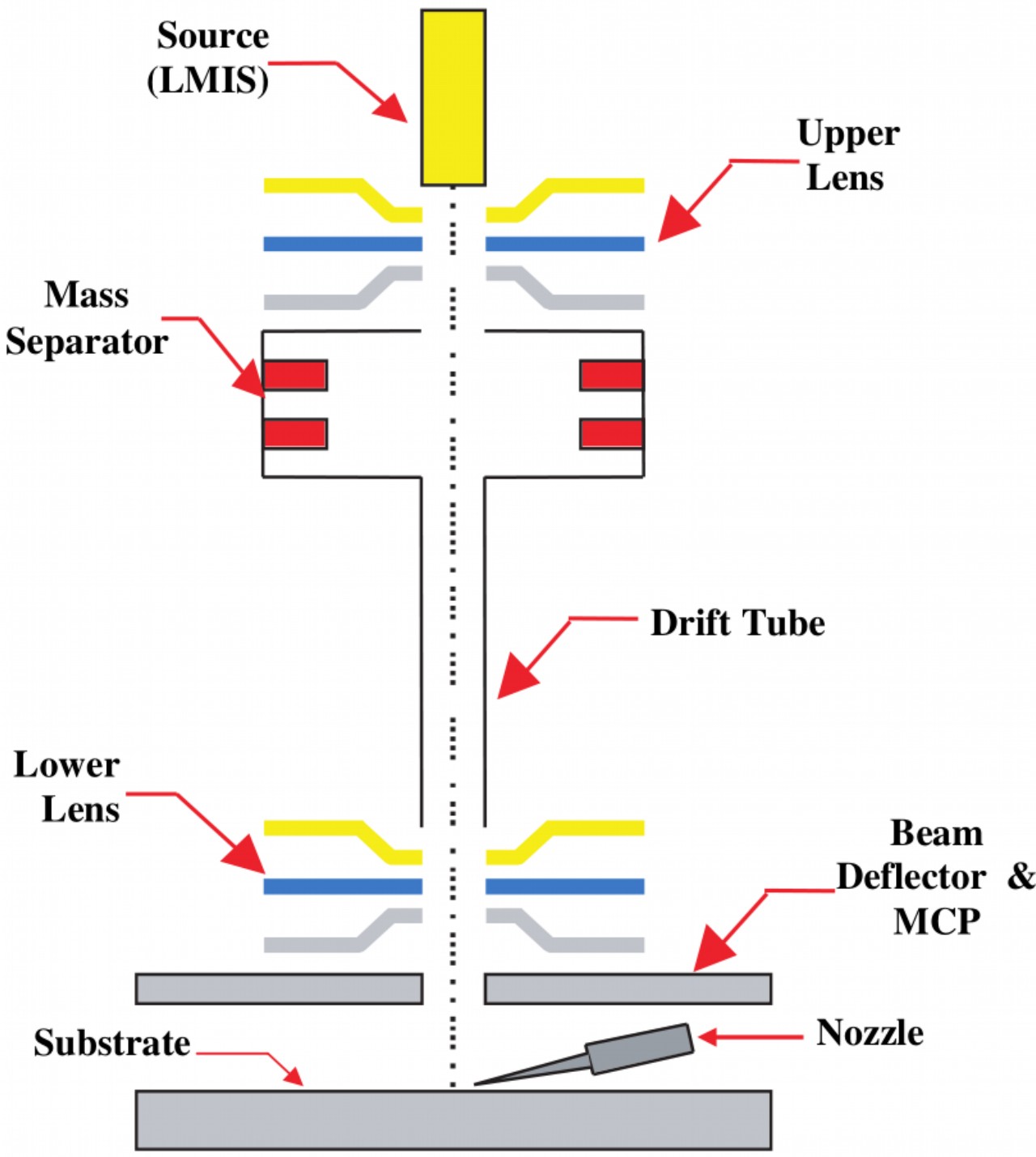}
    \caption{Schematic of an FIB system showing the ion source, collimating lens, mass separator, drift tube, focussing lens, beam deflector, multi-channel plate, nozzle, and target substrate.
    Reproduced with permission from \cite{Tseng_JMM_2004}, all rights reserved, copyright IOP Publishing.} \label{fig_s3_fib}
\end{figure}

An overview of an FIB system is described by Tseng \cite{Tseng_JMM_2004} and a schematic is shown in Fig.~\ref{fig_s3_fib}.
A liquid metal ion source (LMIS) is used to extract positively charged ions, which are then collimated by the upper lens.
The most common type of ion used for milling is gallium.
Next, the ion beam passes through a mass separator, so that only a certain number of ions with a specific mass/charge ratio pass, and a drift tube, to remove ions that are not vertically oriented.
A second lower lens is placed after the drift tube in order to reduce the ion beam spot size and improve the focus.
Finally, the ion beam passes through an electrostatic beam deflector, to control the beam trajectory. A multi-channel plate (MCP) is used to record secondary electron emission.
The system is usually in a vacuum chamber in order to increase the mean-free-path of the ions.
Additionally, a nozzle can be used for FIB-induced deposition.

FIB milling is widely used in the semiconductor industry for high-quality, high-precision fabrication of devices $\leq$1~$\mu$m \cite{Tseng_JMM_2004} and for transmission electron microscope sample preparation \cite{Reyntjens_JMM_2001}.
Several studies have already demonstrated that it is possible to mill structures in micro- and nanofibres for quantum and atom optics applications.
In the following subsections, FIB milled structures on fibres are reviewed.

\subsection{Fibre-based Optical Devices Milled using FIB} \label{fib_structures}

Ion milling for cleaving tapered optical fibres tips was shown as early as 1999 by Fielding et al. \cite{Fielding_JLT_1999}. The authors studied mode evolution by cleaving at different points along the taper and examining the near-field light pattern. 
Around the same time, ion implantation for creating long period gratings in optical fibres was shown \cite{Fujimaki_OL_2000, Fujimaki_JAP_2000, vonBibra_OE_2001}, using FIB milling of structures in optical fibres, followed shortly thereafter by Hodzic et al. \cite{Hodzic_JVS_2003, Hodzic_JLT_2004} and Schiappelli et al. \cite{Schiappelli_ME_2004}.
 
\begin{figure}
    \centering
    \includegraphics[scale=0.6]{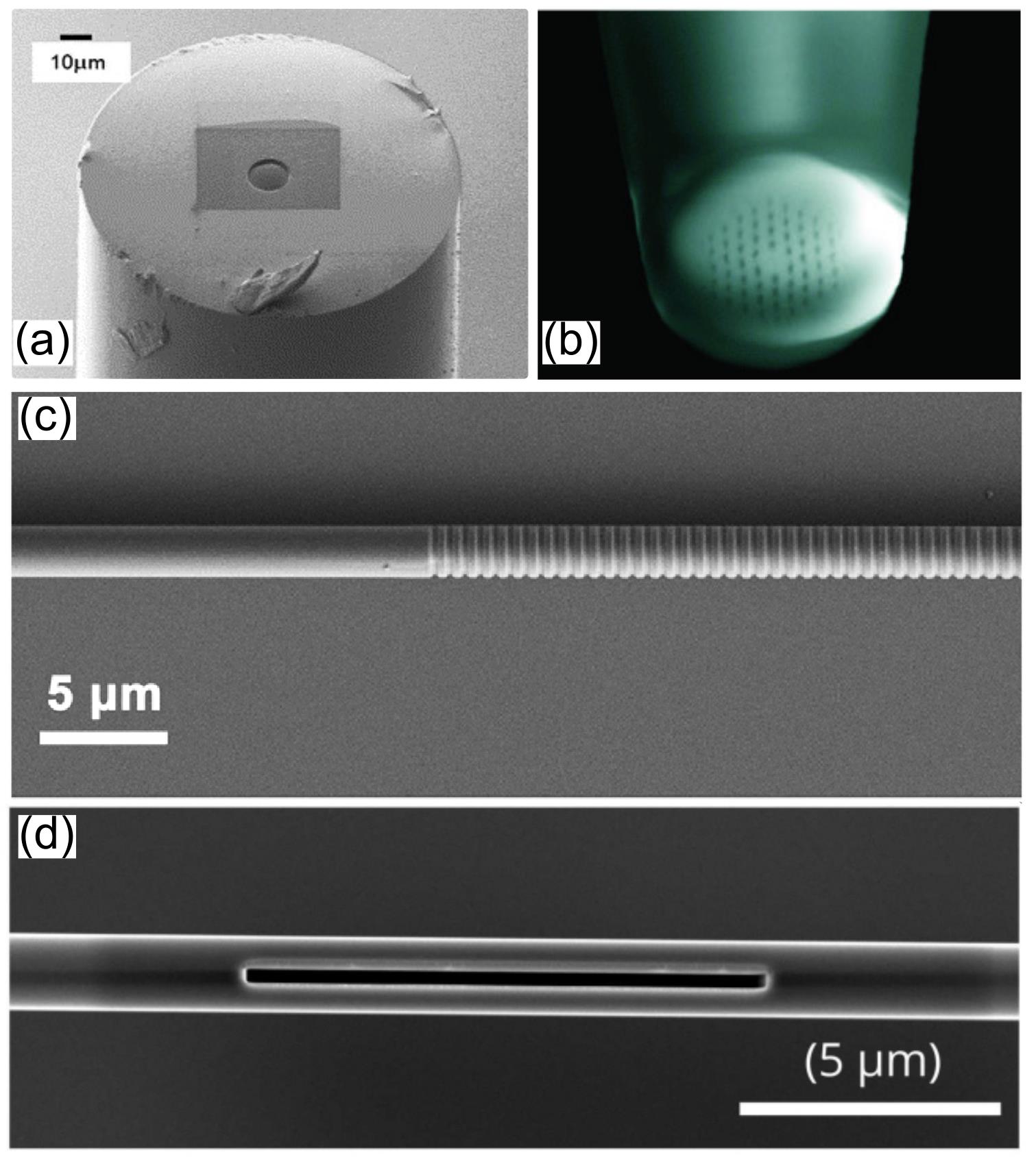}
    \caption{A selection of fibre structures milled using the FIB technique:
    (a) microlens, reprinted from \cite{Schiappelli_ME_2004}, with permission from Elsevier;
    (b) taper cleaving, reprinted with permission from \cite{Gibson_OE_2005}, copyright The Optical Society;
    (c) fibre Bragg gratings, reprinted with permission from \cite{Liu_OL_2011}, copyright The Optical Society;
    (d) slotted fibre, reprinted with permission from \cite{Daly_OE_2016}, copyright The Optical Society.} \label{fig_s3_t01}
\end{figure}

Hodzic et al. \cite{Hodzic_JVS_2003,Hodzic_JLT_2004} milled periodic grooves in the waist of a biconically tapered, single-mode fibre and studied mode coupling.
Grooves, 0.5~$\mu$m wide, 10~$\mu$m long, and separated by 0.5~$\mu$m, were cut into a fibre with a diameter of 35~$\mu$m over a length of 31~$\mu$m. The cuts introduced a refractive index perturbation into the fibre that selectively coupled the modes in the waist region.
Meanwhile, Schiappelli et al. \cite{Schiappelli_ME_2004} fabricated a microlens (shown in Fig.~\ref{fig_s3_t01}a) directly on the cleaved tip of a single-mode fibre in order to improve light coupling from the fibre to a lithium niobate waveguide.
The microlens achieved a coupling efficiency of 67\% for 1550~nm wavelength and was milled by patterning 10 circular crowns with different diameters.
The lens had a focal length of 58.6~$\mu$m and a total diameter of 16~$\mu$m. It was milled directly by  programming the FIB machine after selecting suitable parameters such as beam limiting aperture size, ion dose, dwell time, and beam current.

In the years that followed, several groups demonstrated structures on optical fibres that were fabricated using FIB milling for different applications.
Gibson et al. \cite{Gibson_OE_2005} cleaved tapered photonic crystal fibre (PCF) tips to expose and characterise their  nanohole arrays.
These tips are useful in microfluidic and biophotonic applications and, prior to FIB cleaving techniques, it was not possible to observe nanohole structures in tips with diameters as small as 3.5~$\mu$m (Fig.~\ref{fig_s3_t01}b).
Also for PCFs, it is interesting to access the channels to make microvolume cells for inserting materials such as gases and liquids. 
In this way, Martelli et al. \cite{Martelli_OE_2007} used the FIB technique to mill holes in the sides of PCFs to enable the interconnection of internal microchannels.

Bragg grating structures in microfibres were demonstrated by Liu et al. \cite{Liu_OL_2011}, see Fig.~\ref{fig_s3_t01}c.
The structure was fabricated on a 1.8~$\mu$m diameter taper and had grooves 100~nm deep with a 576~nm pitch. The total length was 518~$\mu$m.
The grooves could achieve high-index contrast by periodic fluctuations between air/silica.
The microfibre grating was then used for refractive index sensing, achieving a sensitivity of 660~nm/RIU (refractive index units) for a wavelength range of 1587-1616~nm.
Also, the authors showed good mechanical strength by bending a grating milled in a 900~nm diameter taper to a radius of 97~$\mu$m.

A slotted fibre device used to trap nanoparticles was demonstrated by Daly et al. \cite{Daly_OE_2016}, see Fig.~\ref{fig_s3_t01}d. 
Slotted waveguides can exhibit symmetric or anti-symmetric modes depending on the phase difference between the upper and lower sections.
The device, originally proposed for trapping cold neutral atoms \cite{Daly_NJP_2014}, was made by milling a 300~nm $\times$ 10~$\mu$m slot into a 1.4~$\mu$m microfibre.
The taper was submerged in water containing 200~nm fluorescing particles, with a density of approximately $10^9$~particles/mL, equivalent to an average particle occupancy of $<$~1 over the volume of the slot.
The authors showed that the device could trap single nanoparticles using an optical power as low as 1.2~mW.
This is important work as it demonstrated trapping at low power, which is essential when working with biological samples or for atoms in ultrahigh vacuum.

For ease of comparison, Table~\ref{tab_s3_01} summarises the main parameters used by the cited works in this subsection.

\subsection{Fibre-Based Optical Cavities Milled with FIB} \label{fib_cavities}

The preceding sections highlighted some of what has been achieved using FIB milling in relatively large fibres or tapered fibres and illustrates a variety of reflecting, focusing, or light guiding geometries and their applications. To date, several studies have demonstrated optical cavities milled directly onto micro- or nanofibres using FIB milling. This section reviews some of these works and their applications.

\begin{figure}
    \centering
    \includegraphics[scale=0.52]{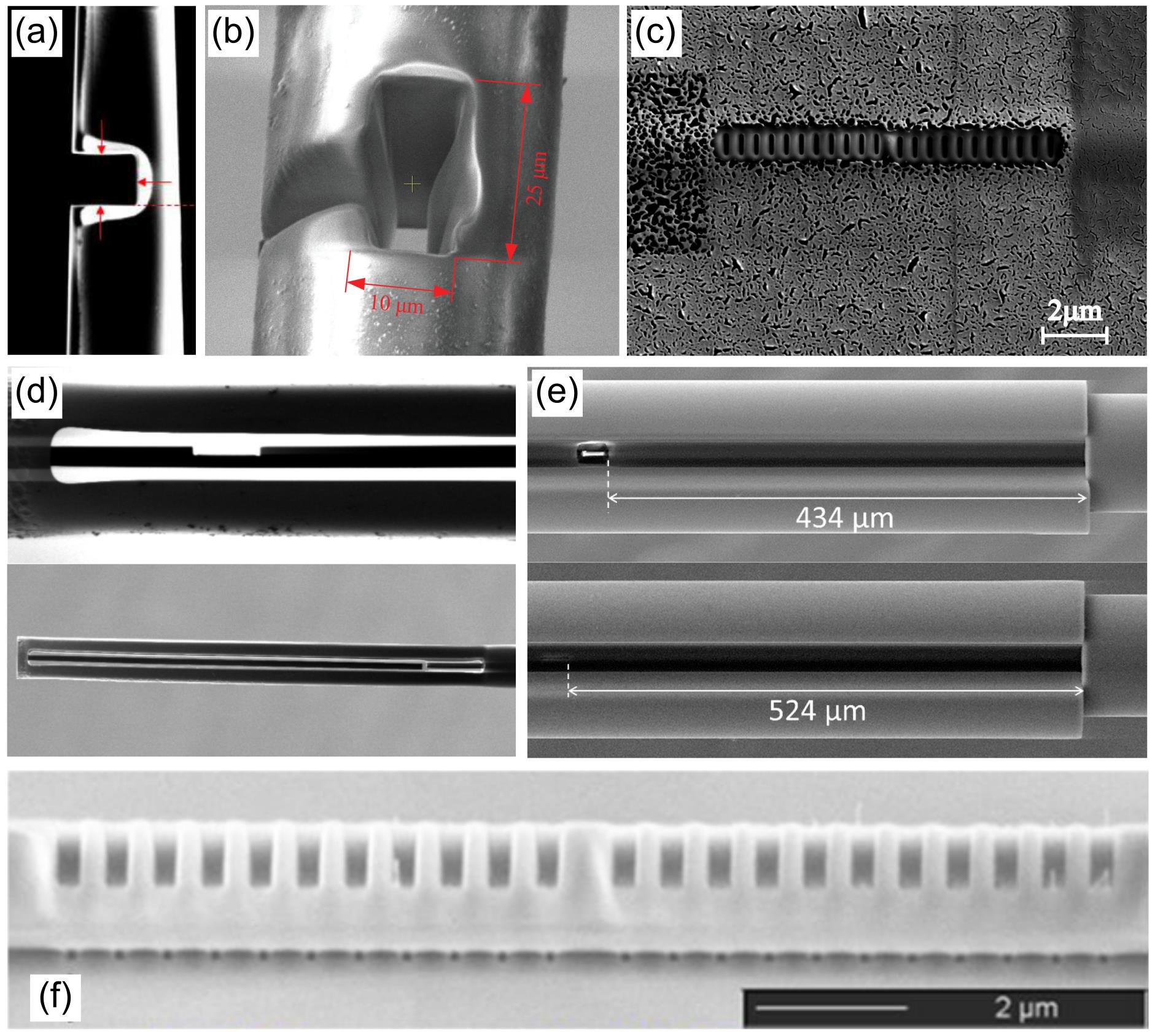}
    \caption{SEM images of microfibre cavities structures demonstrated in the literature, milled using FIB:
    (a) Fabry-P\'{e}rot modal interferometer, reprinted with permission from \cite{Kou_OE_2010}, copyright The Optical Society;
    (b) Fabry-P\'{e}rot refractive index sensor, reprinted from \cite{Yuan_RSI_2011}, with the permission of AIP Publishing;
    (c) Bragg grating, reprinted from \cite{Ding_APL_2011}, with the permission of AIP Publishing;
    (d) indentation (top) and slot cantilever (bottom), reprinted with permission from \cite{Andre_OE_2014}, copyright The Optical Society; 
    (e) penetrating (top) and shallow (bottom) cavities, reprinted with permission from \cite{WarrenSmith_OE_2016}, copyright The Optical Society;
    (f) short (1.02~$\mu$m) photonic crystal cavity, reprinted with permission from \cite{Sun_JPCS_2016}.} \label{fig_s3_t02}
\end{figure}

Kou et al. \cite{Kou_OE_2010} and Yuan et al. \cite{Yuan_RSI_2011} demonstrated microcavities in tapered fibre tips for creating Fabry-P\'{e}rot type refractive index sensors.
Kou et al. \cite{Kou_OE_2010} used their device to demonstrate temperature sensing up to 520$^{\circ}$C with a 4.4~$\mu$m by 5~$\mu$m microcavity structure (see Fig.~\ref{fig_s3_t02}a), and achieved a sensitivity of  $\sim$20~pm/$^{\circ}$C at $\sim$1550~nm.
Yuan et al. \cite{Yuan_RSI_2011} also showed temperature sensing, up to 63.8$^{\circ}$C, with a structure (shown in Fig.~\ref{fig_s3_t02}b)  25~$\mu$m $\times$ 10~$\mu$m and obtained a sensitivity of $\sim$0.5~nm/$^{\circ}$C or equivalently  1731~nm/RIU for a wavelength range of 1520-1620~nm.
Interestingly, \cite{Yuan_RSI_2011} performed FIB milling in the same area twice: first with a large current and high milling rate (referred to in Tab.~\ref{tab_s3_01} as \textit{M}) to shape the structure, and then again at a small current to polish the structure (referred to in Tab.~\ref{tab_s3_01} as \textit{P}).
This method is common to achieve a higher resolution without long milling times \cite{Yuan_RSI_2011,Andre_OE_2014,WarrenSmith_OE_2016} and is useful to minimise charge accumulation on the fibre.

Ding et al. \cite{Ding_APL_2011} fabricated a cavity with phase-shifted Bragg gratings milled into a microfibre with a diameter of 2.3~um.
The microfibre (see Fig.~\ref{fig_s3_t02}c) was partially embedded in polymer to increase its sturdiness and was coated in gold to avoid charging during the ion beam milling (see Section 4).
In analogy to photonic crystal structures, holes of 156~nm $\times$ 718~nm were cut into the fibre at a period of 467~nm to produce a microcavity length of 687~nm.
The gold was removed after FIB milling and a $Q$-factor of 60 was obtained at a resonance wavelength of $\sim$1180~nm.
Although a simple structure, further optimization of the  manufacturing parameters provided a more compact device with improved performance.
The authors predicted that Q-factors in excess of 1000 could be achieved for a larger number of notches.

Andr\'{e} et al. \cite{Andre_OE_2014} demonstrated two Fabry-P\'{e}rot structures for temperature sensing.
The structures were made in a chemically etched microfibre with a 15~$\mu$m diameter.
Before FIB milling, the microfibre were sputter-coated with a thin tantalum film (ca. 50~nm). To form the cavity, a gap was milled in the microwire using FIB.
In the first structure, the milled gap just created an indentation rather than cutting through the entire microwire (see Fig.~\ref{fig_s3_t02}d top). 
This resulted in a low-finesse Fabry-P\'{e}rot (FP) cavity at a wavelength of 1550~nm.
In the second structure, the gap did cut through the microwire, creating a slot cantilever (see Fig.~\ref{fig_s3_t02}d bottom).
This structure also behaved as an FP cavity, but with a higher Q factor, in that the reflecting interfaces were on top of the fibre and the silica-to-air interface at the air gap.
In milling these structures, an ion current of approximately 1~nA was used for a primary coarse milling of the cavities.
After this, polishing was performed using a much smaller current. 
For temperature characterisation, the Fabry-P\'erot structures were placed inside a tubular oven and the temperature was varied from 100 - 550$^{\circ}$C.
For both structures, sensitivities from 11.5~pm/K to 15.5~pm/K, with slightly quadratic slopes, were obtained up to 550$^{\circ}$C.
In addition, the cantilever structure was used for vibration sensing and could detects frequencies from 1~Hz to 40~kHz.

Cavities in an exposed-core PCF were fabricated by Warren-Smith et al. \cite{WarrenSmith_OE_2016}.
Exposed-core PCFs have a portion of the external cladding removed to access the core.
The exposed-core fibre was made from a specially prepared preform that had three holes drilled into the rod in an equilateral triangle configuration with a 0.4~mm separation between the holes.
A slot was cut into one of the holes using a diamond blade, exposing the core.
The preform was then drawn with a positive hole pressure to prevent the holes from collapsing.
The resulting fibre had an outer diameter of 160~$\mu$m and an effective core diameter of 6.8~$\mu$m. It was then spliced to a single-mode fibre.
The authors milled two type of structures into the side of the exposed-core fibre. First, they created a slot that penetrated the exposed core, called a penetrating cavity (see Fig.~\ref{fig_s3_t02}e top), and then they milled slots that did not penetrate, called shallow cavities (Fig.~\ref{fig_s3_t02}e bottom).  For milling, the samples were mounted onto an aluminium block using a silver paste.
A tantalum coating was used to discharge the fibre during ion milling. An ion beam was also used to remove the coating post-fabrication.
Total milling times used were three hours for the larger (penetrating) cavity and 12 minutes for the smaller (non-penetrating) one.
Cavity lengths for the penetrating and shallow cavities were, respectively, 34.6~$\mu$m and 28.9~$\mu$m, and the authors measured the reflection spectra and showed an FSR of, respectively, 4.7~THz and 0.24~THz around 1250 nm.
The results also demonstrated that different cavities can be readily de-multiplexed, allowing for simultaneous measurement of different parameters (e.g. refractive index, temperature, and strain).

Sun et al. \cite{Sun_JPCS_2016} designed and demonstrated a photonic crystal cavity in a microfibre, see Fig.~\ref{fig_s3_t02}f.
Rectangular holes, 1.2~$\mu$m wide, were milled into a 1.7~$\mu$m diameter microfibre to create a cavity with a resonant wavelength of 1530~nm.
Two types of cavities were reported: a long (50~$\mu$m) one with a $Q$ of 800 and a short one (1.02~$\mu$m) with a $Q$ of 73.
Arguably, the most important properties of an optical cavity are its losses and mode volume.
To determine the effect of gallium ion milling on the optical losses, the authors of Ref. \cite{Sun_JPCS_2016} measured the insertion losses for microfibres exposed to different gallium ion doses.
They assumed that ions concentrate on the surface (to a depth of $\sim$50~nm) and partially overlap with the fundamental mode.
The authors showed that the imaginary part of the effective modal index increased with the ion dose, indicating that it may be due to gallium contamination.

The mode volume of fibre-based optical cavities can be extremely small.
It depends on the fibre diameter and the distance between the mirrors, as well as the effective length of the mirror.
Therefore, fibre-based microcavities  can produce mode volumes of only a few $\mu$m$^{3}$. The previous examples demonstrate milling in fibres with diameters of a few microns.
To further reduce the volume and increase transverse confinement, it is necessary to make cavities in fibres with diameters of a few hundred nanometres.
The advantage of ion milling is also clear here since the distance between the mirrors is only limited by the resolution of the ion beam. 
This and aforementioned properties make nanofibres the most suitable structures for creating fibre cavities with dimensions close to the physical limits of confinement.

\begin{figure}
    \centering
    \includegraphics[scale=0.9]{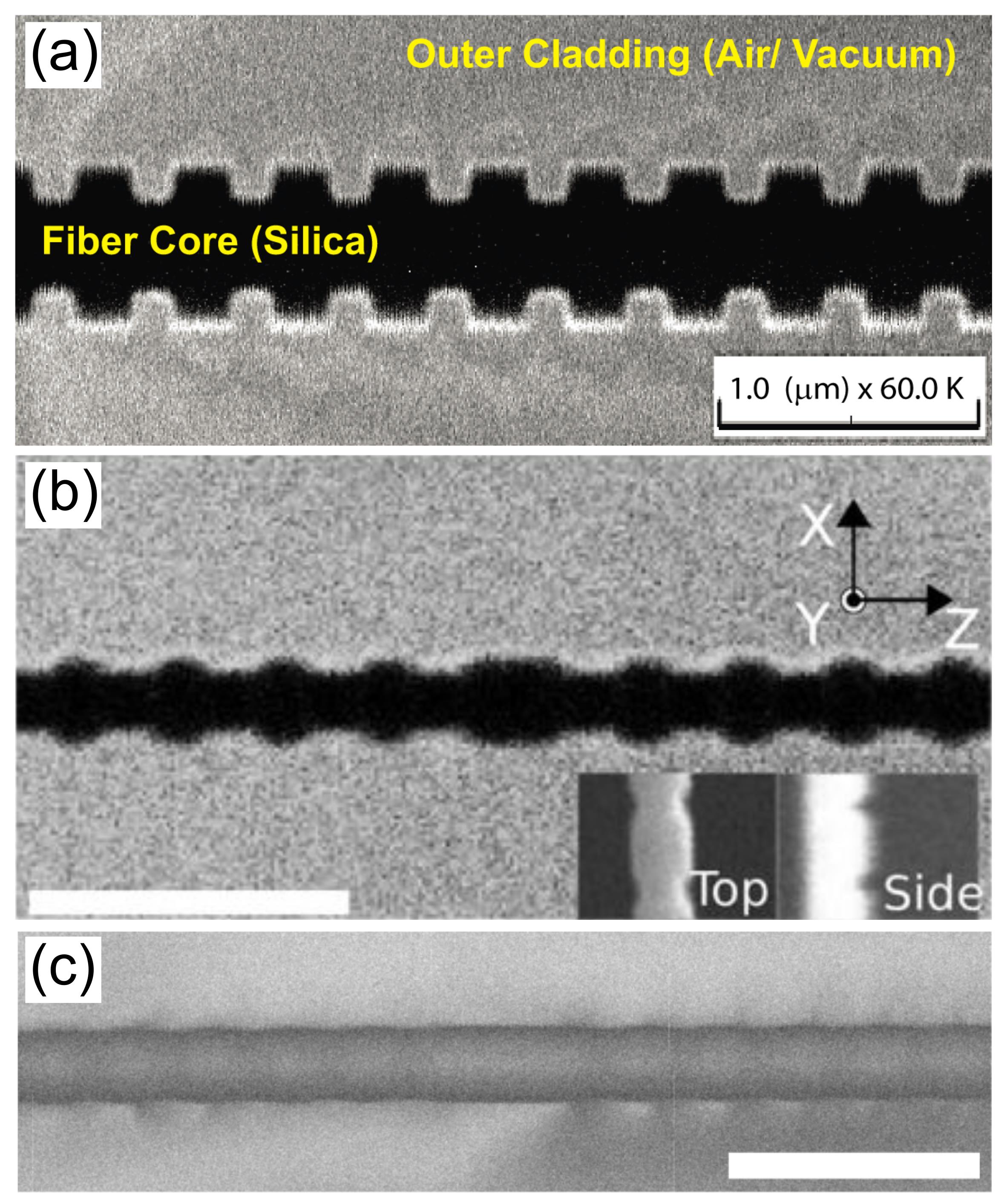}
    \caption{Scanning ion microscope (SIM) images of optical cavities with Bragg grating structures on nanofibres fabricated using FIB.
    (a) $\sim$~560~nm diameter fibre and grooves of $\sim$100~nm deep and $\sim$150~nm wide, reprinted with permission from \cite{Nayak_OE_2011}, copyright The Optical Society;
    (b) scale bar of 1~$\mu$m, $\sim$~270~nm diameter fibre and grooves of 45~nm, 300~nm pitch and 450~nm defect length, reprinted from \cite{Schell_SR_2015} with permission, Creative Commons license;
    (c) scale bar of 1~$\mu$m, $\sim$~310~nm diameter fibre and grooves of 30~nm, 320~nm pitch and 840~nm defect length, reprinted with permission from \cite{Takashima_OE_2019}, copyright The Optical Society.} \label{fig_s3_t03}
\end{figure}

FIB milling of Bragg gratings to generate cavities in nanofibres has been experimentally demonstrated \cite{Nayak_OE_2011} and theoretically studied \cite{LeKien_JMO_2012}.
Nayak et al. \cite{Nayak_OE_2011} milled two types of Bragg grating mirrors to form a fibre cavity. The first had a fixed period, i.e., it was not chirped, and was used to produce short cavities (see Fig.~\ref{fig_s3_t03}a). The second was chirped, i.e., non-periodic mirrors were used to compensate for the resonance wavelength shift caused by diameter fluctuations.
The chirped mirrors were used to form long cavities.
Gratings were made by milling grooves in the fibre $\sim$100~nm deep $\times$ $\sim$150~nm wide.
For the non-chirped mirror, the grating had 120 periods with a pitch of 360~nm.
The chirped mirror was milled with 4 gratings of 60 periods each and with different pitches ranging from 358-365~nm.
The resonance wavelength was around 852~nm and the achieved $F$ was up to 117 with a transmission of 25$\%$ for the short cavity.
For the long cavity, $F$ was 20 with a transmission  up to 20$\%$.
The authors also showed a shorter cavity with a length of 50~$\mu$m and a reduced groove size ($\sim$50~nm depth, $\sim$150~nm width) formed from 180 periods of 345~nm pitch.
Here, they obtained an $F$ of 35 with a transmission of 80$\%$. 
The demonstrated cavities were polarisation-dependent due to the difference between the effective refractive index of x- and y-polarised modes.
Most fibre structures using FIB milling demonstrate this behaviour, because the ion beam orientation is usually along the fibre axis.

A nanofibre cavity composed of a photonic crystal structure was demonstrated by Wuttke \cite{Wuttke_PHD_2014}.
Interestingly, to discharge the sample, the author used a copper mirror in a custom designed holder, placed underneath the nanofibre. The nanofibre was held to the Cu mirror by van der Waals forces.
The copper mirror was grooved in order to avoid redeposition of sputtered copper atoms that were removed unintentionally during milling.
Holes of 136~nm$\times$51~nm were cut into the fibre to produce a grating with a 355~nm pitch and 30 periods.
A high $Q$ of 9.6$\times$10$^{4}$ and a moderate finesse of 11.1$\pm$0.2 were obtained for a resonance wavelength of $\sim$838~nm.
Wuttke also studied the effect of gallium contamination on the cavity losses and demonstrated an annealing procedure for removing contamination and improving quality.
The annealing process involved placing the nanofibre cavity in a vacuum chamber (at 10$^{-5}$~mbar) and directing laser light into the cavity so that heating occurred in areas with high concentrations of gallium absorption.
Gallium ions were then removed by diffusion and evaporation.
The author showed that a heating power of 4~$\mu$W for 2 minutes was sufficient to improve the reflection as well as blue shift the reflection peak by 50~nm.
Increasing the heating power to 100~$\mu$W (for 2 min), the blue reflection peak shifted another 40~nm and the transmission improved from 60\% to 80\%.
This technique is very promising for improving optical characteristics of fibre devices milled using a Ga FIB.

An optical nanofibre  with Bragg grating structures was also demonstrated by Schell et al. \cite{Schell_SR_2015}.
This device, shown in Fig.~\ref{fig_s3_t03}b, was formed by cutting small grooves in the side of the fibre.
The grooves had a period of 160~nm with a 300~nm pitch and a groove depth of 45~nm.
The achieved $Q$-factor was 250, for a cavity resonance wavelength around 630~nm with a corresponding mode volume of only 0.7~$\mu$m$^{3}$.
A key capability of an optical cavity is the tunability of the cavity modes.
Another advantage of nanofibre cavities is their sensitivity to mechanical strain.
In this case the cavity resonance was tuned by applying strain via a piezo-activated translation stage on which the tapered fibre was mounted.
The cavity was tuned at a rate of 0.05~nm/$\mu$m, giving a tuning span up to 25.8~nm.
The authors also demonstrated coupling of quantum dot emission into the cavity. A quantum dot was deposited on the cavity surface using a tungsten tip and emission into the cavity was enhanced by a factor of 3 with respect to free-space. Increased channelling of photons into the cavity was attributed to Purcell enhancement.

\begin{figure}
    \centering
    \includegraphics[scale=0.32]{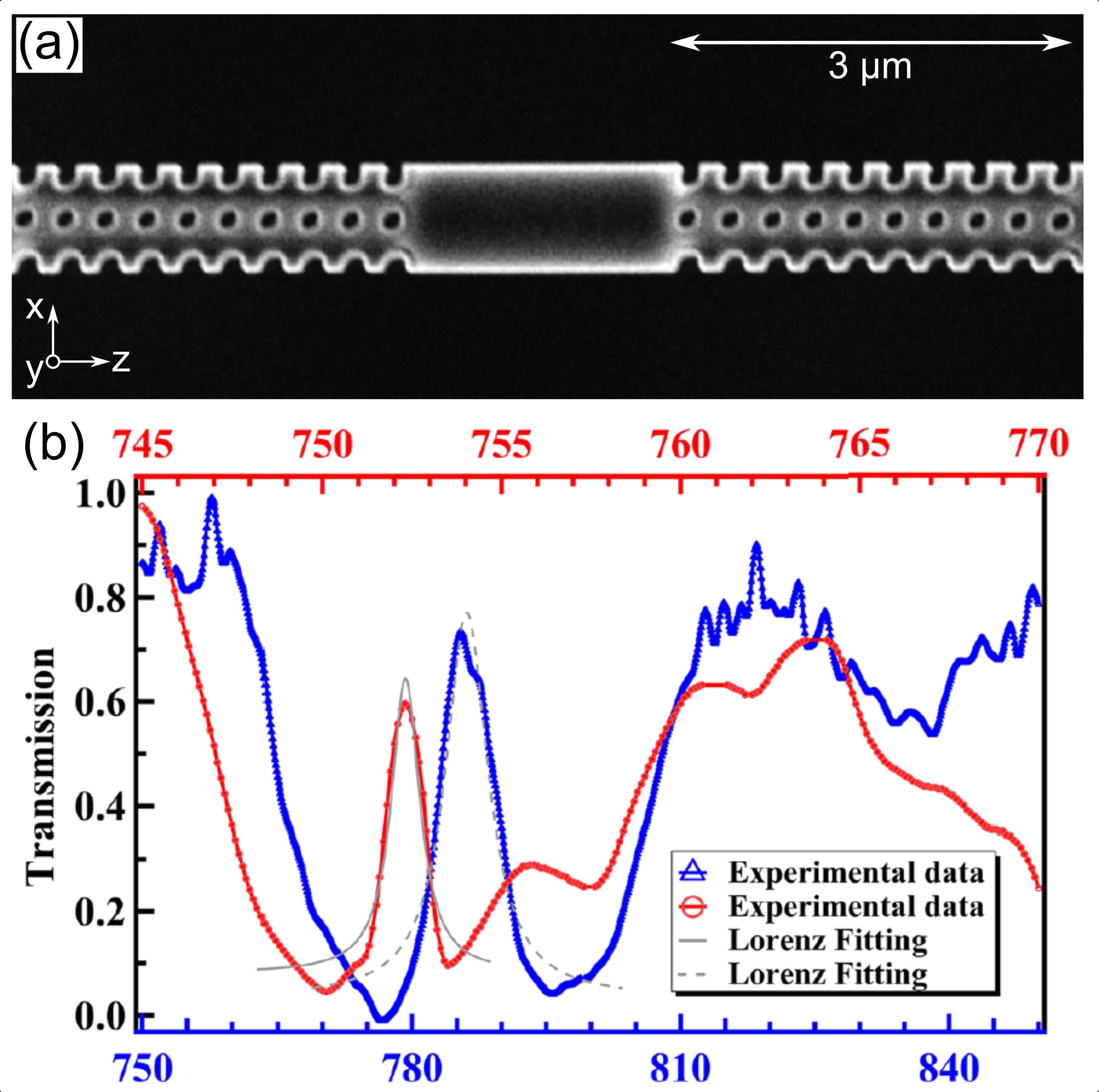}
    \caption{Combined structure of a photonic crystal with a Bragg grating.
    (a) SEM image. The cavity length is 2.2~~$\mu$m.
    (b) Transmission spectra per wavelength (nm units) measured for x- (red) and y- (blue) polarisations.
    Reprinted from \cite{Li_APL_2017} with the permission of AIP Publishing.} \label{fig_s3_fli}
\end{figure}

An alternative nanofibre cavity design that combined a photonic crystal structure with Bragg gratings was proposed by Li et al. \cite{Li_APL_2017}.
The authors showed that the combined structure can achieve higher reflectivity even with a low number of periods.
An SEM picture of the structure is shown in Fig.~\ref{fig_s3_fli}a.
Milled holes were $\sim$100~nm$\times$100~nm with a 310~nm pitch.
Finite-difference time-domain (FDTD) simulation of the structure predicted a reflectivity higher than 80$\%$ with only 30 periods, resulting in a very compact device.
Experimentally, a structure with 20 periods was fabricated using FIB milling. The normalised cavity transmission is shown in Fig.~\ref{fig_s3_fli}b for x- (red) and y- (blue) polarisations.
The resonance wavelength was around 780~nm. The mode volume was $\sim$1.05~$\mu$m$^{3}$ and a $Q$-factor up to 784$\pm$87 was achieved.
The following year, the authors extended their study to include a slot inside the cavity between the mirrors \cite{Li_OL_2018}, similar to what was shown in \cite{Daly_OE_2016}.
The slot permits deposition of  single emitters at the position of strongest field intensity, that is, within the slot, thereby enhancing emission into the cavity.
The authors estimated a 10$\times$ increase in the Purcell factor compared to the case in which a single emitter is placed on the fibre surface outside the slot.

As discussed earlier,  deposition of gallium ions can be a source of loss in the fibre cavity.
To overcome this limitation, ion milling using a different ion was demonstrated by Takashima et al. \cite{Takashima_OE_2019}.
A helium ion beam was used to mill Bragg gratings and to generate a nanofibre cavity with higher $Q$ (Fig.~\ref{fig_s3_t03}c) than the same structure milled with gallium FIB.
Additionally, the authors reported that gallium ions have poorer accuracy and offer lower resolution than helium ions.
Using a He FIB, they obtained a resolution down to 1~nm, compared to only tens of nm for a Ga FIB.
In addition, He is a noble gas; therefore, electrical, optical, and chemical alteration of the target material is minimised \cite{Ward_JVSTB_2006}, i.e., it can be considered contamination free.
In this work \cite{Takashima_OE_2019}, the authors compared cavities produced with He and Ga ions. They showed that the $Q$-factor was higher in the cavity milled with He (450 versus 250 for the Ga FIB milling).
The He FIB was used to mill a cavity with a high $Q$-factor of 4170 and a wavelength resonance peak at 699.80~nm.
The Bragg mirrors had groove depths of 30~nm, with a pitch of 320~nm, and 640 periods.
This significant increase in resolution and $Q$-factor opens up possibilities for further refinement of optical cavities fabricated with FIB milling and demonstrates that this technique will continue to contribute to development of high-quality optical devices.

As a quick reference for the reader, Table~\ref{tab_s3_02} and Table~\ref{tab_s3_03} contain summaries of the important parameters of devices discussed in this section.  

\afterpage{
\begin{table*}
\tiny

\caption{Summary of fibre structures by FIB milling.} \label{tab_s3_01}
\setlength{\tabcolsep}{0.4cm}
\begin{tabular*}{\linewidth}{c c c c c c l l}
\hline\hline\noalign{\smallskip}
\textbf{Ref.} & \textbf{LMIS} & \textbf{V} & \textbf{I} & \textbf{Spot Size} & $\mathbf{d_{Fibre}}$ & \textbf{Fibre Structure} & \textbf{Discharge}\\
\noalign{\smallskip}\hline\hline\noalign{\smallskip}
\cite{Hodzic_JVS_2003,Hodzic_JLT_2004} & Ga & 30~kV & 300~pA & 100~nm & 35~$\mu$m & Grooves in biconically tapers & Al coating ($\sim$50~nm-thick)\\
\noalign{\smallskip}\hline\noalign{\smallskip}
\cite{Schiappelli_ME_2004} & Ga & 30~kV & - & 7~nm & - & Microlens on fibre tip & - \\
\noalign{\smallskip}\hline\noalign{\smallskip}
\cite{Gibson_OE_2005} & Ga & 30~kV & 280~pA & 500~nm & - & Taper cleaving & - \\
\noalign{\smallskip}\hline\noalign{\smallskip}
\cite{Martelli_OE_2007} & Ga & - & 20~nA & $\sim$500~nm & $\sim$100~$\mu$m & Holes to interconnect fibre & Conductive carbon tape \\
\noalign{\smallskip}\hline\noalign{\smallskip}
\cite{Liu_OL_2011} & Ga & 30~kV & 70~pA & - & 1.8~$\mu$m & Microfibre Bragg grating & Si conductive substrate \\
\noalign{\smallskip}\hline\noalign{\smallskip}
\cite{Daly_OE_2016} & Ga & - & - & - & 1.4~$\mu$m & Sloted microfibre & ITO coating ($\sim$5~nm-thick) \\
\noalign{\smallskip}\hline\hline
\end{tabular*}
\smallskip\smallskip

\caption{Summary of microfibre cavities produced with FIB milling. In the current column, \textit{I}, \textit{M} and \textit{P} stand for, respectively, the current values used for milling and polishing.} \label{tab_s3_02}
\setlength{\tabcolsep}{0.11cm}
\begin{tabular*}{\linewidth}{c c c c c c c c c l l}
\hline\hline\noalign{\smallskip}
\textbf{Ref.} & \textbf{LMIS} & \textbf{V} & \textbf{I} & \textbf{Spot Size} & $\mathbf{d_{Fibre}}$ & $\mathbf{L_{Cavity}}$ & $\mathbf{Q}$ & $\mathbf{F}$ & \textbf{Cavity Structure} & \textbf{Discharge}\\
\noalign{\smallskip}\hline\hline\noalign{\smallskip}
\cite{Kou_OE_2010} & Ga & 30~kV & 291~pA & 55~nm & 9.2~$\mu$m & - & - & - & Taper tip & Al coating (150~nm-thick) \\
\noalign{\smallskip}\hline\noalign{\smallskip}
\cite{Yuan_RSI_2011} & Ga & - & 20~nA(M)/3~nA(P) & 300-500~nm(M)/10~nm(P) & 32~$\mu$m & - & - & - & Taper tip & Conductive carbon tape \\
\noalign{\smallskip}\hline\noalign{\smallskip}
\cite{Ding_APL_2011} & Ga & 30~kV & 93~pA & $\leq$30~nm & $\sim$2.3~$\mu$m & 687.5~nm & 60 & - & Bragg grating  & Au coating (50~nm-thick) \\
\noalign{\smallskip}\hline\noalign{\smallskip}
\cite{Andre_OE_2014} & Ga & - & 1~nA(M)/100-300~pA(P) & - & 15~$\mu$m &  &  &  & Etched microwire & Ta coating (50~nm-thick) \\
 &  &  &  &  &  & $\sim$172~$\mu$m & - & - & Indentation &  \\
 &  &  &  &  &  & $\sim$1026~$\mu$m & - & - & Slot cantilever &  \\
\noalign{\smallskip}\hline\noalign{\smallskip}
\cite{WarrenSmith_OE_2016} & - & - &  & - & - &  &  &  & Exposed-core fibre & Ta coating (50~nm-thick) \\
 &  &  & 2~nA(M)/600~pA(P) &  &  & 34.6~$\mu$m & - & - & Penetrating cavity &  \\
 &  &  & 630~pA &  &  & 28.9~$\mu$m & - & - & Shallow cavity &  \\
\noalign{\smallskip}\hline\noalign{\smallskip}
\cite{Sun_JPCS_2016} & Ga & - & - & - & 1.7~$\mu$m &  &  &  & Photonic crystal & Si conductive substrate \\
 &  &  &  &  &  & 50~$\mu$m & $\sim$800 & $\sim$7.5 & Long cavity & \\
 &  &  &  &  &  & 1.02~$\mu$m & 73 & - & Short cavity & \\
\noalign{\smallskip}\hline\hline
\end{tabular*}
\smallskip\smallskip

\caption{Summary of nanofibre cavities by FIB milling.} \label{tab_s3_03}
\setlength{\tabcolsep}{0.11cm}
\begin{tabular*}{\linewidth}{c c c c c c c c c l l}
\hline\hline\noalign{\smallskip}
\textbf{Ref.} & \textbf{LMIS} & \textbf{V} & \textbf{I} & \textbf{Spot Size} & $\mathbf{d_{Fibre}}$ & $\mathbf{L_{Cavity}}$ & $\mathbf{Q}$ & $\mathbf{F}$ & \textbf{Cavity Structure} & \textbf{Discharge}\\
\noalign{\smallskip}\hline\hline\noalign{\smallskip}
\cite{Nayak_OE_2011} & Ga & 30~kV & $\sim$10~pA & $\sim$14~nm &  &  &  &  & Bragg grating & - \\
 &  &  &  &  & $\sim$560~nm & 100~$\mu$m & - & $\sim$42-117 & Short cavity, single-period &  \\
 &  &  &  &  & $\sim$560~nm & $\sim$5~mm & - & 20 & Long cavity, chirped &  \\
 &  &  &  &  & $\sim$520~nm & 50~$\mu$m & - & 35 & Short cavity, single-period, reduced groove &  \\
\noalign{\smallskip}\hline\noalign{\smallskip}
\cite{Wuttke_PHD_2014} & Ga & 30~kV & 1-10~pA & $\sim$10~nm & $\sim$520~nm & 2.39$\pm$0.1~mm & 9.6$\times$10$^{4}$ & 11.1$\pm$0.2 & Photonic crystal & Cu conductive substrate \\
\noalign{\smallskip}\hline\noalign{\smallskip}
\cite{Schell_SR_2015} & Ga & 30~kV & 9.3~pA & 13~nm & 270~nm & 450~nm & 250 & - & Bragg grating & - \\
\noalign{\smallskip}\hline\noalign{\smallskip}
\cite{Li_APL_2017} & Ga & 30~kV & 7~pA & 9~nm & $\sim$830~nm & $\sim$2.2~$\mu$m & 784$\pm$87 & - & Photonic crystal and Bragg grating & ITO coating \\
\noalign{\smallskip}\hline\noalign{\smallskip}
\cite{Takashima_OE_2019} & He & - & 1~pA & - & 306~nm & 840~nm & 4170 & - & Bragg grating & - \\
\noalign{\smallskip}\hline\hline
\end{tabular*}

\end{table*}}
\section{Experimental Details for Fabrication of a Nanofibre-Based Optical Cavity with FIB Milling} \label{experiment}

A nanofibre cavity with a complex mirror structure was previously fabricated by our group \cite{Li_APL_2017}, as discussed in Section \ref{fib_cavities}. To achieve high-quality devices, one must be careful about the preparation, fabrication, and handling of tapered fibres, both pre- and post-milling.
Moreover, we found that a number of additional steps were needed to achieve stability during milling.
In this section, we discuss our findings.
The initial fabrication, in \cite{Li_APL_2017}, involved a three-step process:
(i) the optical fibre was tapered and then fixed onto an aluminium substrate, using optical glue.
In this case, the tapered fibre was suspended over the substrate; 
(ii) the sample/tapered fibre was coated with an ITO layer to reduce surface charges;
(iii) the sample was milled with a Ga source FIB.
However, fabrication results using this process were not consistent, primarily due to the low conductivity of the thin ($\sim$20~nm) ITO layer and vibrations of the silica fibre when impacted by Ga ions.  Therefore, we sought a way to dispense with this fibre coating step in order to improve repeatability and sample quality.

The versatility of ion beam milling enables direct fabrication of structures on optical nanofibres and other devices. Low lateral scattering of ions ensures that only the intended regions are exposed to the ion beam \cite{Tseng_S_2005}.
However, this works only if charge accumulation is minimal, which can be achieved by discharging the sample.
The usual practices for solving this issue, shown in the column \textit{Discharge} in Tab.~\ref{tab_s3_01}-\ref{tab_s3_03}, include direct coating of the sample with a conductive material (that may be removed later) \cite{Hodzic_JVS_2003, Hodzic_JLT_2004, Kou_OE_2010, Ding_APL_2011, Andre_OE_2014, WarrenSmith_OE_2016, Daly_OE_2016, Li_APL_2017} or placing it on top of a conductive substrate \cite{Martelli_OE_2007, Yuan_RSI_2011, Liu_OL_2011, Wuttke_PHD_2014, Sun_JPCS_2016}.
Negatively charging the sample with an electron beam from an SEM can also be used to avoid this charging effect and can  partially compensate for sample charging \cite{Martelli_OE_2007}.

For micro- or nanofibres that are to be used for experiments involving quantum emitters (e.g. quantum dots or neutral atoms) interacting with light in the fibre's evanescent field, fibre coatings, such as the conductive ITO layer, can lead to undesirable absorption and scattering losses.
Also, coating the nanofibre before milling necessitates an additional procedure, which involves  handling the nanofibre in a sputtering chamber. Additional handling increases the difficulty of the process and exposes the fibre to different sources of contamination.
As shown in Sections \ref{fib_structures} and \ref{fib_cavities}, it may be possible to remove the coating by etching or even milling again using an FIB, but such invasive methods can damage the final structure.

If a fibre is uncoated and a conductive substrate is used, to avoid contamination due to re-deposition of the sputtered atoms during milling, the substrate material should be chosen so that it does not absorb significantly in the wavelength range being used in later experiments.
This is important when milling holes through the fibre because the ion beam will also cut into the substrate, and sputtered atoms from the substrate can re-deposit on the holes, causing optical losses in cavity transmission, as demonstrated previously for a copper substrate \cite{Wuttke_PHD_2014}.

 Recently, we developed techniques to overcome the obstacles described above by simply using a Si slab coated with ITO as a conductive substrate. ITO is almost transparent for visible wavelengths and Maniscalco et al. \cite{Maniscalco_TSF_2014} showed that a 238~nm ITO layer sputtered on soda lime glass can have transmission values exceeding 75$\%$ for a wavelength range of 600-800~nm.
As a result, contamination can be minimised when using an ITO substrate instead of a metal substrate. This enables us to eliminate the conductive coating on the nanofibre, reducing loss due to handling and increasing mechanical stability of the nanofibre during milling. In this section, the sequence for fabrication of a nanofibre-based cavity is described in detail.

\subsection{Optical Nanofibre Preparation} \label{exp_fab}

Optical nanofibres are produced using a flame brushing technique, described in detail elsewhere \cite{Ward_RSI_2014} and illustrated in Fig.~\ref{fig_s2_fnanofibre}.
We use a hydrogen/oxygen flame as a heat source and routinely produce nanofibres with diameters as small as 400~nm and transmission as high as 95\%.
For this work, a commercial SM800 fibre from Fibercore, with an initial cladding diameter of 125~$\mu$m, was tapered down to $\sim$800~nm with a transmission of 98$\%$ at 780~nm.
The fibre cavity resonance wavelength was chosen to be near 780~nm, to coincide with the emission profile of a particular CdSeTe quantum dot.

\begin{figure}
    \centering
    \includegraphics[scale=0.41]{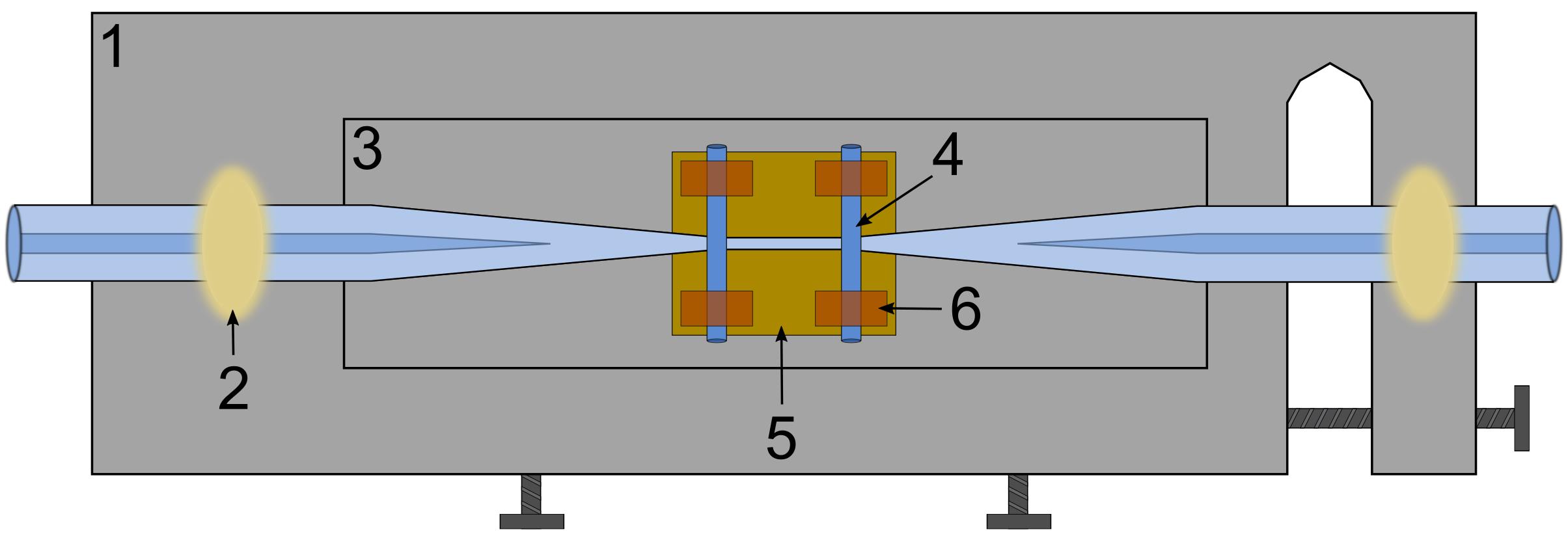}
    \caption{Schematic of the base plate with an ITO-coated Si substrate for FIB milling. The numbers in the figure corresponds to: the aluminium base plate (1), UV optical glue (2), a removable section (3), a holding fibre (4), the ITO-coated Si substrate (5), and Kapton tape (6).} \label{fig_s4_fsub}
\end{figure}

After tapering, the fibre was glued onto an aluminium base plate that was specifically designed to hold the tapered fibre during FIB milling.
A schematic of the base plate is shown in Fig.~\ref{fig_s4_fsub}.
The base plate has a rectangular slot, in which a rectangular block sits.
Stuck on top of the block is an ITO coated Si substrate.
The block is held in place in the slot by screws.
The height of the substrate relative to the tapered fibre can be adjusted prior to tightening the screws.
The main idea of the aluminium base plate is to ensure that the nanofibre is physically in contact with the ITO-coated Si substrate.

The Si substrate was sputtered with ITO in a sputtering deposition machine to create a uniform ITO coating. We sputter in argon plasma, using a flow of 20 sccm, at 100$^{\circ}$C, with 100~W of power, and rotation at 5~rpm.
Exposure time of the substrate to the ITO target was typically 480~s, yielding a layer thickness of $\sim$40~nm.
The substrate was fixed to the block with carbon tape and was designed to be $\sim$0.5~mm higher than the base plate and approximately the length of the tapered region. This way it is in contact with the tapered waist length. To doubly ensure that the taper waist is in contact with the substrate, two small pieces of optical fibre (cladding diameter of 80~$\mu$m), which we term "holding" fibres, were used to push the taper downward. These were attached to the substrate using Kapton tape.
These small pieces of fibres are coated with Pt-Pd to reduce the surface charges.
Note that the tapered optical fibre is quite flexible and can be easily pushed into contact with the ITO-coated substrate using the holding fibres; however, one should be careful not to over-tighten the tapered fibre during fabrication. In fact, it is best to leave the fibre slight slack.
Removing the substrate from beneath the tapered fibre post-milling is also straightforward. It is released by simply removing the Kapton tape along with the holding fibres and then removing the ITO-coated Si substrate by loosening the side bolts. 
Additionally, we include a movable ``arm" on the mount so that we can elongate the tapered fibre to remove slack and to shift the resonance wavelength of the milled fibre cavity (see Fig.~\ref{fig_s4_fsub}).

\subsection{FIB Milling of the Optical Nanofibre} \label{exp_fib}

Despite advantages of using a He ion source \cite{Takashima_OE_2019}, Ga is still the most accessible ion type in FIB systems. The system we use is a gallium ion source FIB integrated with an SEM (FEI system, model Helios NanoLab G3 UC), in which the ion beam is tilted at 52$^{\circ}$ in relation to the electron beam.
In our experiments, the Ga ion beam was extracted using a voltage of 30~kV.
The beam dwell time was  1~$\mu$s, corresponding to the time the beam spends at each pixel.
The sputter rate is characterised as the target volume removed per charge and is defined by the material. For silica, its value is 0.24~$\mu$m$^{3}$/nC at 30~kV \cite{FEI_manual_2014}.
The beam current was selected between 2-7~pA, corresponding to beam diameters ranging from  $\sim$7-9.8~nm.
The current value was chosen for each sample according to its conductivity.
During milling, we noticed that if the nanofibre is in good contact with the ITO-coated substrate, the conductivity is such that we can operate at a current as low as 2~pA.
However, if certain areas of the nanofibre do not maintain good contact with the ITO, the conductivity decreases and the current needs to be increased to 7~pA.
For aligning the ion beam before milling the nanofibre, a portion of the ITO-coated Si substrate was used to adjust the focus and astigmatism of the electrostatic lens of the FIB.
Alignment of the ion beam requires extremely sensitive control of the focus and astigmatism and, unlike for an electron beam, even a short period of ion beam exposure can significantly damage the nanofibre.
Therefore, at high magnification, we first mill holes on the substrate, the same size and pitch as the intended structure, without exposing the nanofibre to the ion beam.
Milled structures are imaged by the SEM and this procedure is repeated until structures are clearly defined. Once the beam alignment is optimised, the magnification is decreased until it is possible to see the nanofibre and we start to mill the nanofibre without changing the focus or astigmatism. 

\begin{figure}
    \centering
    \includegraphics[scale=0.31]{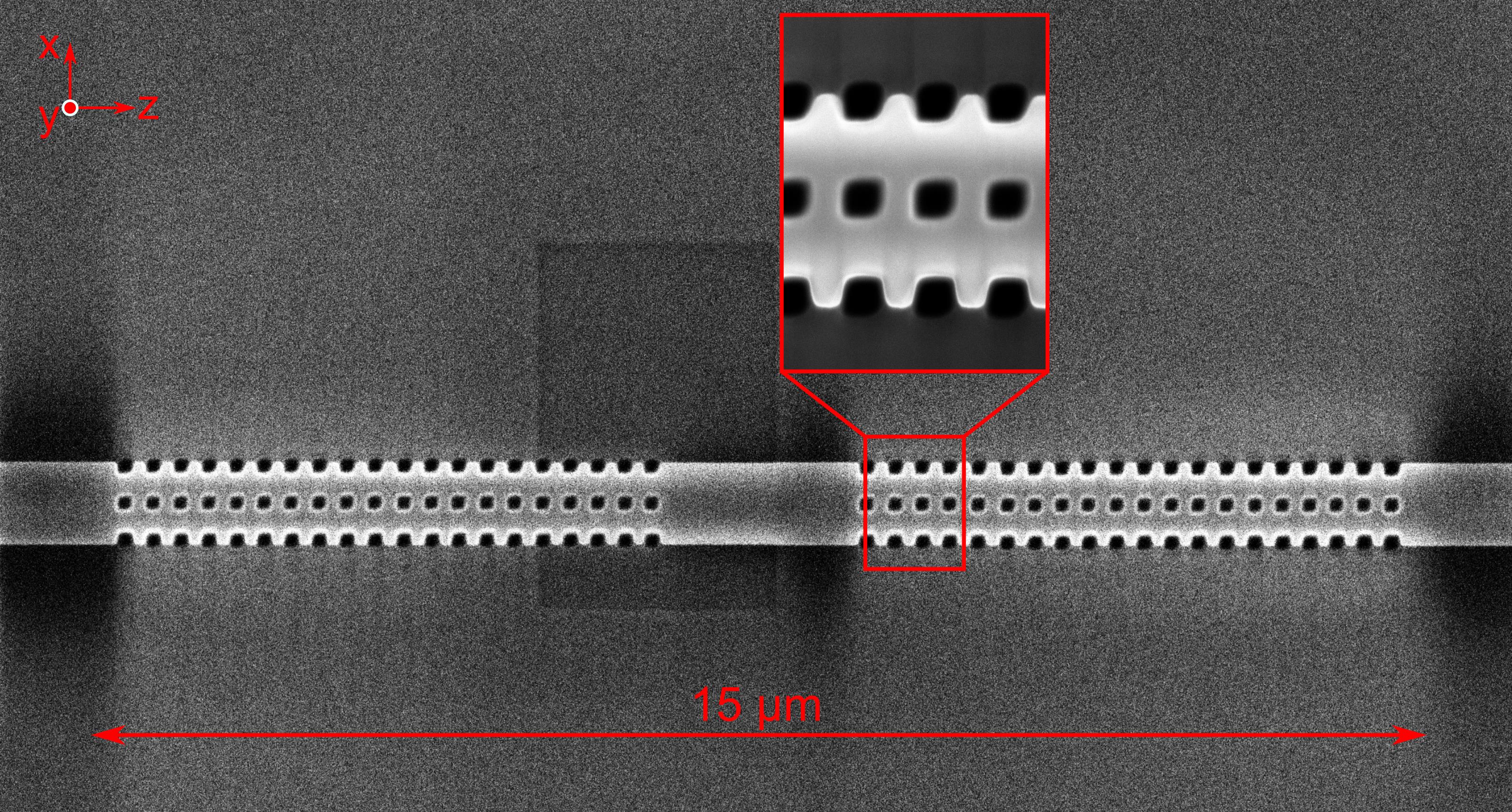}
    \caption{SEM image of a 970-nm diameter optical nanofibre milled using a Ga FIB.
    The structure size measured was $\sim$141.6~nm$\times$130.4~nm with a pitch of $\sim$326.3~nm.} \label{fig_s4_fsample}
\end{figure}

For the sake of completeness, here we discuss milling a device similar to that of our earlier work \cite{Li_APL_2017}.
This triplex air-cube structure combines Bragg and photonic crystal mirrors, see Section \ref{fib_cavities}, and can have high reflectivity for a small number of structures. 
In this work, we defined the triplex air-cube to consist of 120~nm square holes, with 21 periods for each mirror, and a pitch of 320~nm.
SEM images of the milled cavities are shown in Fig.~\ref{fig_s4_fsample}. The nanofibre diameter was 970~nm and the measured size of the air-cube structures was $\sim$141.6~nm$\times$130.4~nm (x-axis$\times$z-axis, see Fig.~\ref{fig_s4_fsample}) with a pitch of $\sim$326.3~nm.
The measured cavity length was 2.2~$\mu$m.
An error between defined and measured values was observed to be $\sim$21.6~nm$\times$10.4~nm for hole size and $\sim$6.3~nm for pitch.
This can be attributed to the beam stability and resolution.
The nanofibre surface is curved and, in practice, resolution tends to be lower for milling this type of surface.
Also, we noticed that the error in hole size along the x-axis is larger than along the z-axis. The reason for this is that the ion beam is scanned in the x-axis in a serpentine pattern, i. e., the scan direction switches to the opposite after each column is scanned.
Therefore, the x-axis of the nanofibre is more susceptible to vibrations caused by the ion beam scan. 
It is important to note that the overall shape and pitch of the structure is maintained and errors on the order of a few to tens of nm are reasonable for high precision milling of complex structures.
We have milled a number of nanofibre cavities using the ITO-coated Si substrate and found that the overall stability and conductivity lead to improved repeatability and reduced turn-around time. 

Nanofibre-based cavities fabricated by the process described here are small, integrated devices suitable for studying single quantum emitters in fibre-based quantum networks. Optical characterisation of this device can be found in \cite{Li_APL_2017}.

\section{Conclusions and Outlook} \label{conclusions}

The main advantages of working with nanofibres and nanofibre cavities for cavity quantum electrodynamics have been presented. 
We described  Fabry-Per\'ot type optical cavities in which mirrors are fabricated directly on the waist of a tapered optical fibre. These inline fibre cavities can easily be integrated into conventional optical systems, giving them a unique advantage over other designs.
Different fibre cavity fabrication techniques have been reported in the literature, so we compiled and briefly reviewed each work here to give the reader an  overview of the state-of-the-art.
Since the FIB enables high precision and high-quality milling for micro- and nanostructures, we conclude that this is still one of the most suitable techniques for  producing high-quality tapered fibre-based optical cavities with complex designs, capable of achieving both strong and weak coupling regimes.

A summary of the literature shows that FIB milling allows for the creation of different types of devices directly in optical fibres, such as lenses, sensing devices (e.g. microresonators in tapered tips and Bragg gratings), and even slotted fibres that can trap nanoparticles. After a brief overview of the FIB milling technique itself, we presented some of the applications of these optical devices which were milled into optical fibres.

FIB milling with He ions was shown to have several advantages over Ga, such as higher resolution and without contamination generated by the process.
However, Ga is still the most accessible type of FIB and although it can lead to contaminated milled structures, this problem can be minimised with an annealing procedure after milling.
Thus, Ga ion milling still remains a promising method for future fabrication of devices with very high optical quality.

Lastly, our customised fabrication process using Ga FIB milling to obtain nanofibre-based cavities was described.
A specially designed base plate with an ITO-coated Si substrate was demonstrated to control mechanical instability generated by charge accumulation.
Even though we use a Ga ion beam, which has lower resolution than He, we have shown  that good methodology can improve the quality of the milling process and structures of a few tens of nm in size can be achieved in nanofibres with errors of a few nm.
We believe that the fabrication technique described here facilitates fabrication of complex mirror structures in nanofibres with a high level of repeatability; thus, it is especially useful for quantum optics applications aimed at integrated single quantum emitters.

\begin{acknowledgements}
This work was supported by the Okinawa Institute of Science and Technology Graduate University. The authors acknowledge W. Li and J. Keloth for very useful discussions and advice, S. D. Aird for reviewing the manuscript, technical assistance from K. Karlsson, M. Ozer and the Nanofabrication Section in OIST, and general research support from E. Nakamura.
\end{acknowledgements}

\bibliographystyle{spphys.bst}
\bibliography{References.bib}

\begin{thebibliography}{100}
\providecommand{\url}[1]{{#1}}
\providecommand{\urlprefix}{URL }
\expandafter\ifx\csname urlstyle\endcsname\relax
  \providecommand{\doi}[1]{DOI \discretionary{}{}{}#1}\else
  \providecommand{\doi}{DOI \discretionary{}{}{}\begingroup
  \urlstyle{rm}\Url}\fi

\bibitem{Birks_JLT_1992}
T.A. {Birks}, Y.W. {Li}, J. Lightwave Technol. \textbf{10}(4), 432 (1992)

\bibitem{Tong_N_2003}
L.~Tong, R.R. Gattassand, J.B. Ashcom, S.~He, J.~Lou, M.~Shen, I.~Maxwell,
  E.~Mazur, Nature \textbf{426}, 816 (2003)

\bibitem{Tong_OE_2004}
L.~Tong, J.~Lou, E.~Mazur, Opt. Express \textbf{12}(6), 1025 (2004)

\bibitem{Brambilla_AOP_2009}
G.~Brambilla, F.~Xu, P.~Horak, Y.~Jung, F.~Koizumi, N.P. Sessions,
  E.~Koukharenko, X.~Feng, G.S. Murugan, J.S. Wilkinson, D.J. Richardson, Adv.
  Opt. Photon. \textbf{1}(1), 107 (2009)

\bibitem{Tong_Sumetsky_2010}
L.~Tong, M.~Sumetsky, \emph{Subwavelength and Nanometer Diameter Optical
  Fibers} (Springer Heidelberg Dordrecht London New York, 2010)

\bibitem{Villatoro_OE_2005}
J.~Villatoro, D.~{Monz\'{o}n-Hern\'{a}ndez}, Opt. Express \textbf{13}(13), 5087
  (2005)

\bibitem{Lou_S_2014}
J.~Lou, Y.~Wang, L.~Tong, Sensors \textbf{14}(4), 5823 (2014)

\bibitem{Beugnot_NC_2014}
J.C. Beugnot, S.~Lebrun, G.~Pauliat, H.~Maillotte, V.~Laude, T.~Sylvestre, Nat.
  Commun. \textbf{5}(1), 5242 (2014)

\bibitem{Brambilla_OFT_2010}
G.~Brambilla, Opt. Fiber Technol. \textbf{16}(6), 331  (2010)

\bibitem{Tong_S_2018}
L.~Tong, Sensors \textbf{18}(3), 903 (2018)

\bibitem{Sague_PRL_2007}
G.~Sagu\'e, E.~Vetsch, W.~Alt, D.~Meschede, A.~Rauschenbeutel, Phys. Rev. Lett.
  \textbf{99}, 163602 (2007)

\bibitem{Vetsch_PRL_2010}
E.~Vetsch, D.~Reitz, G.~Sagu\'e, R.~Schmidt, S.T. Dawkins, A.~Rauschenbeutel,
  Phys. Rev. Lett. \textbf{104}, 203603 (2010)

\bibitem{Hendrickson_PRL_2010}
S.M. Hendrickson, M.M. Lai, T.B. Pittman, J.D. Franson, Phys. Rev. Lett.
  \textbf{105}, 173602 (2010)

\bibitem{Jones_JOSAB_2014}
D.E. Jones, J.D. Franson, T.B. Pittman, J. Opt. Soc. Am. B \textbf{31}(8), 1997
  (2014)

\bibitem{Kien_PRA_2017}
F.L. Kien, A.~Rauschenbeutel, Phys. Rev. A \textbf{95}, 023838 (2017)

\bibitem{Kien_PRA_2018}
F.L. Kien, T.~Ray, T.~Nieddu, T.~Busch, S.~{Nic Chormaic}, Phys. Rev. A
  \textbf{97}, 013821 (2018)

\bibitem{Aoki_N_2006}
T.~Aoki, B.~Dayan, E.~Wilcut, W.P. Bowen, A.S. Parkins, T.J. Kippenberg, K.J.
  Vahala, H.J. Kimble, Nature \textbf{443}(7112), 671 (2006)

\bibitem{Shoppova_APL_2007}
S.I. Shopova, H.~Zhou, X.~Fan, P.~Zhang, Appl. Phys. Lett. \textbf{90}(22),
  221101 (2007)

\bibitem{Ward_LPR_2011}
J.~Ward, O.~Benson, Laser Photonics Rev. \textbf{5}(4), 553 (2011)

\bibitem{Lei_OE_2017}
F.~Lei, Y.~Yang, J.M. Ward, S.~{Nic Chormaic}, Opt. Express \textbf{25}(20),
  24679 (2017)

\bibitem{Nayak_NJP_2008}
K.P. Nayak, K.~Hakuta, New J. Phys. \textbf{10}(5), 053003 (2008)

\bibitem{Ruddell_O_2017}
S.K. Ruddell, K.E. Webb, I.~Herrera, A.S. Parkins, M.D. Hoogerland, Optica
  \textbf{4}(5), 576 (2017)

\bibitem{Solano_Adv_2017}
P.~Solano, J.A. Grover, J.E. Hoffman, S.~Ravets, F.K. Fatemi, L.A. Orozco, S.L.
  Rolston,  (Academic Press, 2017), pp. 439--505

\bibitem{Kato_NC_2019}
S.~Kato, N.~N{\'e}met, K.~Senga, S.~Mizukami, X.~Huang, S.~Parkins, T.~Aoki,
  Nat. Commun. \textbf{10}(1), 1160 (2019)

\bibitem{Araneda_NP_2019}
G.~Araneda, S.~Walser, Y.~Colombe, D.B. Higginbottom, J.~Volz, R.~Blatt,
  A.~Rauschenbeutel, Nat. Phys. \textbf{15}(1), 17 (2019)

\bibitem{Haroche_PT_1989}
S.~Haroche, D.~Kleppner, Phys. Today \textbf{42}, 24 (1989)

\bibitem{Kimble_IOP_1998}
H.J. Kimble, Phys. Scripta \textbf{T76}(1), 127 (1998)

\bibitem{Haroche_CQE_1995}
S.~Haroche, D.~Kleppner, \emph{Cavity Quantum Electrodynamics} (Springer US,
  Boston, MA, 1995), pp. 849--855

\bibitem{Mabuchi_S_2002}
H.~Mabuchi, A.C. Doherty, Science \textbf{298}(5597), 1372 (2002)

\bibitem{Vahala_N_2003}
K.J. Vahala, Nature \textbf{424}(6950), 839 (2003)

\bibitem{Kimble_N_2008}
H.J. Kimble, Nature \textbf{453}(7198), 1023 (2008)

\bibitem{Reiserer_RMP_2015}
A.~Reiserer, G.~Rempe, Rev. Mod. Phys. \textbf{87}, 1379 (2015)

\bibitem{Kien_PRA_2009}
F.L. Kien, K.~Hakuta, Phys. Rev. A \textbf{80}, 053826 (2009)

\bibitem{Nayak_OE_2011}
K.P. Nayak, F.L. Kien, Y.~Kawai, K.~Hakuta, K.~Nakajima, H.T. Miyazaki,
  Y.~Sugimoto, Opt. Express \textbf{19}(15), 14040 (2011)

\bibitem{Wuttke_OL_2012}
C.~Wuttke, M.~Becker, S.~Br\"{u}ckner, M.~Rothhardt, A.~Rauschenbeutel, Opt.
  Lett. \textbf{37}(11), 1949 (2012)

\bibitem{Nayak_OE_2013}
K.P. Nayak, K.~Hakuta, Opt. Express \textbf{21}(2), 2480 (2013)

\bibitem{Yalla_PRL_2014}
R.~Yalla, M.~Sadgrove, K.P. Nayak, K.~Hakuta, Phys. Rev. Lett. \textbf{113},
  143601 (2014)

\bibitem{Daly_SPIE_2015}
M.~Daly, V.G. Truong, S.~{Nic Chormaic}, in \emph{Optical Trapping and Optical
  Micromanipulation XII}, vol. 9548, ed. by K.~Dholakia, G.C. Spalding.
  International Society for Optics and Photonics (SPIE, 2015), vol. 9548, pp.
  202--205

\bibitem{Kato_PRL_2015}
S.~Kato, T.~Aoki, Phys. Rev. Lett. \textbf{115}, 093603 (2015)

\bibitem{Nayak_JO_2018}
K.P. Nayak, M.~Sadgrove, R.~Yalla, F.L. Kien, K.~Hakuta, J. Opt.
  \textbf{20}(7), 073001 (2018)

\bibitem{Nayak_OL_2014}
K.P. Nayak, P.~Zhang, K.~Hakuta, Opt. Lett. \textbf{39}(2), 232 (2014)

\bibitem{Daly_OE_2016}
M.~Daly, V.G. Truong, S.~{Nic Chormaic}, Opt. Express \textbf{24}(13), 14470
  (2016)

\bibitem{Daly_PHD_2017}
M.~Daly, Light-induced interactions using optical near-field devices.
\newblock Ph.D. thesis, Okinawa Institute of Science and Technology Graduate
  University (2017)

\bibitem{Keloth_OL_2017}
J.~Keloth, K.P. Nayak, K.~Hakuta, Opt. Lett. \textbf{42}(5), 1003 (2017)

\bibitem{Li_APL_2017}
W.~Li, J.~Du, V.G. Truong, S.~{Nic Chormaic}, Appl. Phys. Lett.
  \textbf{110}(25), 253102 (2017)

\bibitem{Schell_SR_2015}
A.W. Schell, H.~Takashima, S.~Kamioka, Y.~Oe, M.~Fujiwara, O.~Benson,
  S.~Takeuchi, Sci. Rep. \textbf{5}, 9619 (2015).
\newblock Creative Commons license: http://creativecommons.org/licenses/by/4.0/

\bibitem{Takashima_OE_2019}
H.~Takashima, A.~Fukuda, H.~Maruya, T.~Tashima, A.W. Schell, S.~Takeuchi, Opt.
  Express \textbf{27}(5), 6792 (2019)

\bibitem{Yalla_PRL_2012}
R.~Yalla, F.L. Kien, M.~Morinaga, K.~Hakuta, Phys. Rev. Lett. \textbf{109},
  063602 (2012)

\bibitem{Nayak_OE_2007}
K.P. Nayak, P.N. Melentiev, M.~Morinaga, F.L. Kien, V.I. Balykin, K.~Hakuta,
  Opt. Express \textbf{15}(9), 5431 (2007)

\bibitem{Ward_RSI_2006}
J.M. Ward, D.G. O'Shea, B.J. Shortt, M.J. Morrissey, K.~Deasy, S.~{Nic
  Chormaic}, Rev. Sci. Instrum. \textbf{77}(8), 083105 (2006)

\bibitem{Ward_RSI_2014}
J.M. Ward, A.~Maimaiti, V.H. Le, S.~{Nic Chormaic}, Rev. Sci. Instrum.
  \textbf{85}(11), 111501 (2014)

\bibitem{Kien_OC_2004}
F.L. Kien, J.Q. Liang, K.~Hakuta, V.I. Balykin, Opt. Commun. \textbf{242}(4),
  445 (2004)

\bibitem{Patnaik_PRA_2002}
A.K. Patnaik, J.Q. Liang, K.~Hakuta, Phys. Rev. A \textbf{66}, 063808 (2002)

\bibitem{LeonSaval_OE_2004}
S.G. Leon-Saval, T.A. Birks, W.J. Wadsworth, P.S.J. Russell, M.W. Mason, Opt.
  Express \textbf{12}(13), 2864 (2004)

\bibitem{Kumar_NJP_2015}
R.~Kumar, V.~Gokhroo, S.~{Nic Chormaic}, New J. Phys. \textbf{17}(12), 123012
  (2015)

\bibitem{Morrissey_RSI_2009}
M.J. Morrissey, K.~Deasy, Y.~Wu, S.~Chakrabarti, S.~{Nic Chormaic}, Rev. Sci.
  Instrum. \textbf{80}(5), 053102 (2009)

\bibitem{Brambilla_OL_2007}
G.~Brambilla, G.S. Murugan, J.S. Wilkinson, D.J. Richardson, Opt. Lett.
  \textbf{32}(20), 3041 (2007)

\bibitem{Skelton_JQSRT_2012}
S.~Skelton, M.~Sergides, R.~Patel, E.~Karczewska, O.~Maragó, P.~Jones, J.
  Quant. Spectrosc. Ra. \textbf{113}(18), 2512 (2012).
\newblock Electromagnetic and Light Scattering by non-spherical particles XIII

\bibitem{Daly_NJP_2014}
M.~Daly, V.G. Truong, C.F. Phelan, K.~Deasy, S.~{Nic Chormaic}, New J. Phys.
  \textbf{16}(5), 053052 (2014)

\bibitem{Brambilla_JO_2010}
G.~Brambilla, J. of Opt. \textbf{12}(4), 043001 (2010)

\bibitem{Tong_OC_2012}
L.~Tong, F.~Zi, X.~Guo, J.~Lou, Opt. Commun. \textbf{285}(23), 4641 (2012)

\bibitem{Morrissey_S_2013}
M.J. Morrissey, K.~Deasy, M.~Frawley, R.~Kumar, E.~Prel, L.~Russell, V.G.
  Truong, S.~{Nic Chormaic}, Sensors \textbf{13}(8), 10449 (2013)

\bibitem{Nieddu_JO_2016}
T.~Nieddu, V.~Gokhroo, S.~{Nic Chormaic}, J. Opt. \textbf{18}(5), 053001 (2016)

\bibitem{Brambilla_OE_2004}
G.~Brambilla, V.~Finazzi, D.J. Richardson, Opt. Express \textbf{12}(10), 2258
  (2004)

\bibitem{Clohessy_EL_2005}
A.M. {Clohessy}, N.~{Healy}, D.F. {Murphy}, C.D. {Hussey}, Electron. Lett.
  \textbf{41}(17), 954 (2005)

\bibitem{Lee_CAP_2019}
D.~Lee, K.J. Lee, J.H. Kim, K.~Park, D.~Lee, Y.H. Kim, H.~Shin, Curr. Appl.
  Phys. \textbf{19}(12), 1334 (2019)

\bibitem{Sumetsky_OE_2004}
M.~Sumetsky, Y.~Dulashko, A.~Hale, Opt. Express \textbf{12}(15), 3521 (2004)

\bibitem{Magi_OE_2007}
E.C. M\"{a}gi, L.B. Fu, H.C. Nguyen, M.R.E. Lamont, D.I. Yeom, B.J. Eggleton,
  Opt. Express \textbf{15}(16), 10324 (2007).
\newblock \doi{10.1364/OE.15.010324}

\bibitem{Brambilla_EL_2005}
G.~Brambilla, F.~Koizumi, X.~Feng, D.J. Richardson, Electron. Lett.
  \textbf{41}(7), 400 (2005)

\bibitem{Petcu-Colan_JNOPM_2011}
A.~Petcu-Colan, M.~Frawley, S.~{Nic Chormaic}, J. Nonlinear Opt. Phys.
  \textbf{20}(03), 293 (2011)

\bibitem{Ravets_OE_2013}
S.~Ravets, J.E. Hoffman, L.A. Orozco, S.L. Rolston, G.~Beadie, F.K. Fatemi,
  Opt. Express \textbf{21}(15), 18325 (2013)

\bibitem{Snyder_Love_1983}
A.W. Snyder, J.D. Love, \emph{Optical Waveguide Theory} (Chapman and Hall,
  1983)

\bibitem{Kien_PRA_2005}
F.L. Kien, S.D. Gupta, V.I. Balykin, K.~Hakuta, Phys. Rev. A \textbf{72},
  032509 (2005)

\bibitem{Srinivasan_APL_2007}
K.~Srinivasan, O.~Painter, A.~Stintz, S.~Krishna, Appl. Phys. Lett.
  \textbf{91}(9), 091102 (2007)

\bibitem{Fujiwara_NL_2011}
M.~Fujiwara, K.~Toubaru, T.~Noda, H.Q. Zhao, S.~Takeuchi, Nano Lett.
  \textbf{11}(10), 4362 (2011)

\bibitem{Schroder_OE_2012}
T.~Schr\"{o}der, M.~Fujiwara, T.~Noda, H.Q. Zhao, O.~Benson, S.~Takeuchi, Opt.
  Express \textbf{20}(10), 10490 (2012)

\bibitem{Liebermeister_APL_2014}
L.~Liebermeister, F.~Petersen, A.~v.~M\"{u}nchow, D.~Burchardt,
  J.~Hermelbracht, T.~Tashima, A.W. Schell, O.~Benson, T.~Meinhardt,
  A.~Krueger, A.~Stiebeiner, A.~Rauschenbeutel, H.~Weinfurter, M.~Weber, Appl.
  Phys. Lett. \textbf{104}(3), 031101 (2014)

\bibitem{Schell_ACSP_2017}
A.W. Schell, H.~Takashima, T.T. Tran, I.~Aharonovich, S.~Takeuchi, ACS
  Photonics \textbf{4}(4), 761 (2017)

\bibitem{Hood_PRL_1998}
C.J. Hood, M.S. Chapman, T.W. Lynn, H.J. Kimble, Phys. Rev. Lett. \textbf{80},
  4157 (1998)

\bibitem{Yamamoto_SLI_1999}
Y.~Yamamoto, S.~Inoue, G.~Björk, H.~Heitmann, F.~Matinaga, in
  \emph{Semiconductor Lasers I}, ed. by E.~Kapon, Optics and Photonics
  (Academic Press, San Diego, 1999), pp. 361--441

\bibitem{Gerry_Knight_QO_2005}
C.~Gerry, P.~Knight, \emph{Introductory Quantum Optics} (Cambridge University
  Press, 2005)

\bibitem{Fox_QO_2006}
M.~Fox, \emph{Quantum Optics: An Introduction} (Oxford University Press, 2006)

\bibitem{Purcell_PR_1946}
E.M. Purcell, Phys. Rev. \textbf{69}, 681 (1946)

\bibitem{Reyntjens_JMM_2001}
S.~Reyntjens, R.~Puers, J. Micromech. Microeng. \textbf{11}(4), 287 (2001)

\bibitem{Tseng_JMM_2004}
A.A. Tseng, J. Micromech. Microeng. \textbf{14}(4), R15 (2004)

\bibitem{Tseng_S_2005}
A.A. Tseng, Small \textbf{1}(10), 924 (2005)

\bibitem{Keskinbora_FIB_2019}
K.~Keskinbora, \emph{Prototyping Micro- and Nano-Optics with Focused Ion Beam
  Lithography} (Society of Photo-Optical Instrumentation Engineers (SPIE),
  2019)

\bibitem{Martelli_OE_2007}
C.~Martelli, P.~Olivero, J.~Canning, N.~Groothoff, B.~Gibson, S.~Huntington,
  Opt. Lett. \textbf{32}(11), 1575 (2007)

\bibitem{Sigmund_PR_1969}
P.~Sigmund, Phys. Rev. \textbf{184}, 383 (1969)

\bibitem{Fielding_JLT_1999}
A.J. {Fielding}, K.~{Edinger}, C.C. {Davis}, J. Lightwave Technol.
  \textbf{17}(9), 1649 (1999)

\bibitem{Fujimaki_OL_2000}
M.~Fujimaki, Y.~Ohki, J.L. Brebner, S.~Roorda, Opt. Lett. \textbf{25}(2), 88
  (2000)

\bibitem{Fujimaki_JAP_2000}
M.~Fujimaki, Y.~Nishihara, Y.~Ohki, J.L. Brebner, S.~Roorda, J. Appl. Phys.
  \textbf{88}(10), 5534 (2000)

\bibitem{vonBibra_OE_2001}
M.L. von Bibra, A.~Roberts, J.~Canning, Opt. Lett. \textbf{26}(11), 765 (2001)

\bibitem{Hodzic_JVS_2003}
V.~Hodzic, J.~Orloff, C.~Davis, J. Vac. Sci. Technol. B \textbf{21}(6), 2711
  (2003)

\bibitem{Hodzic_JLT_2004}
V.~Hodzic, J.~Orloff, C.C. Davis, J. Lightwave Technol. \textbf{22}(6), 1610
  (2004)

\bibitem{Schiappelli_ME_2004}
F.~Schiappelli, R.~Kumar, M.~Prasciolu, D.~Cojoc, S.~Cabrini, M.D. Vittorio,
  G.~Visimberga, A.~Gerardino, V.~Degiorgio, E.D. Fabrizio, Microelectron. Eng.
  \textbf{73-74}, 397 (2004)

\bibitem{Gibson_OE_2005}
B.C. Gibson, S.T. Huntington, S.~Rubanov, P.~Olivero,
  K.~Digweed-Lyytik\"{a}inen, J.~Canning, J.D. Love, Opt. Express
  \textbf{13}(22), 9023 (2005)

\bibitem{Liu_OL_2011}
Y.~Liu, C.~Meng, A.P. Zhang, Y.~Xiao, H.~Yu, L.~Tong, Opt. Lett.
  \textbf{36}(16), 3115 (2011)

\bibitem{Kou_OE_2010}
J.L. Kou, J.~Feng, L.~Ye, F.~Xu, Y.Q. Lu, Opt. Express \textbf{18}(13), 14245
  (2010)

\bibitem{Yuan_RSI_2011}
W.~Yuan, F.~Wang, A.~Savenko, D.H. Petersen, O.~Bang, Rev. Sci. Instrum.
  \textbf{82}(7), 076103 (2011)

\bibitem{Andre_OE_2014}
R.M. Andr\'{e}, S.~Pevec, M.~Becker, J.~Dellith, M.~Rothhardt, M.B. Marques,
  D.~Donlagic, H.~Bartelt, O.~Fraz{\~a}o, Opt. Express \textbf{22}(11), 13102
  (2014)

\bibitem{WarrenSmith_OE_2016}
S.C. Warren-Smith, R.M. Andr\'{e}, C.~Perrella, J.~Dellith, H.~Bartelt, Opt.
  Express \textbf{24}(1), 378 (2016)

\bibitem{Ding_APL_2011}
M.~Ding, P.~Wang, T.~Lee, G.~Brambilla, Appl. Phys. Lett. \textbf{99}(5),
  051105 (2011)

\bibitem{Sun_JPCS_2016}
Y.Z. Sun, Y.~Yu, H.L. Liu, Z.Y. Li, W.~Ding, J. Phys. Conf. Ser. \textbf{680},
  012029 (2016)

\bibitem{LeKien_JMO_2012}
F.L. Kien, K.P. Nayak, K.~Hakuta, J. Mod. Optic. \textbf{59}(3), 274 (2012)

\bibitem{Wuttke_PHD_2014}
C.~Wuttke, Thermal excitations of optical nanofibers measured with a
  fiber-integrated {Fabry-P\'erot} cavity.
\newblock Ph.D. thesis, Faculty of Physics at the Johannes Gutenberg
  Unversit\"at Mainz (2014)

\bibitem{Li_OL_2018}
W.~Li, J.~Du, S.~{Nic Chormaic}, Opt. Lett. \textbf{43}(8), 1674 (2018)

\bibitem{Ward_JVSTB_2006}
B.W. Ward, J.A. Notte, N.P. Economou, Journal of Vac. Sci. Technol. B
  \textbf{24}(6), 2871 (2006)

\bibitem{Maniscalco_TSF_2014}
B.~Maniscalco, P.M. Kaminski, J.M. Walls, Thin Solid Films \textbf{550}, 10
  (2014)

\bibitem{FEI_manual_2014}
FEI Company, 5350 NE Dawson Creek Drive, Hillsboro, OR 97124, \emph{Helios
  {NanoLab} {G3UC/G3CX/660} - User Operation Manual} (2014)

\end{thebibliography}

\end{document}